\documentclass[10pt,conference]{IEEEtran}

\usepackage[utf8]{inputenc}
\usepackage[T1]{fontenc}    
\usepackage{hyperref}       
\usepackage{url}            
\usepackage{booktabs}       
\usepackage{amsfonts}       
\usepackage{nicefrac}       
\usepackage{microtype}      

\usepackage{graphicx}
\usepackage{doi}
\usepackage{amsfonts}
\usepackage{xcolor}
\usepackage{physics}
\usepackage{makecell}
\usepackage{multirow}
\usepackage{graphicx}
\usepackage{subcaption}
\usepackage{enumitem}

\usepackage{physics}
\usepackage{algorithm}
\usepackage{algpseudocode}

\usepackage{mathtools} 
\usepackage{array}
\usepackage{wrapfig} 
\usepackage{svg}
\usepackage[capitalise]{cleveref}

\begin{document}

\title{A Hyperparameter Study for Quantum Kernel Methods}

\author{ Sebastian Egginger\IEEEauthorrefmark{1}\IEEEauthorrefmark{2}, Alona Sakhnenko\IEEEauthorrefmark{1}, Jeanette Miriam Lorenz\IEEEauthorrefmark{1}\IEEEauthorrefmark{2} \\
	\IEEEauthorblockA{\IEEEauthorrefmark{1}Fraunhofer Institute for Cognitive Systems IKS,  Munich, Germany}
	\IEEEauthorblockA{\IEEEauthorrefmark{2}Ludwig-Maximilian University, Munich, Germany}
	\texttt{\{sebastian.egginger, alona.sakhnenko jeanette.miriam.lorenz\}@iks.fraunhofer.de} \\
}

\maketitle
\begin{abstract}
Quantum kernel methods are a promising method in quantum machine learning thanks to the guarantees connected to them. Their accessibility for analytic considerations also opens up the possibility of prescreening datasets based on their potential for a quantum advantage. To do so, earlier works developed the geometric difference, which can be understood as a closeness measure between two kernel-based machine learning approaches, most importantly between a quantum kernel and a classical kernel. This metric links the quantum and classical model complexities, and it was developed to bound generalization error. Therefore, it raises the question of how this metric behaves in an empirical setting. In this work, we investigate the effects of hyperparameter choice on the model performance and the generalization gap between classical and quantum kernels. The importance of hyperparameters is well known also for classical machine learning. Of special interest are hyperparameters associated with the quantum Hamiltonian evolution feature map, as well as the number of qubits to trace out before computing a projected quantum kernel. We conduct a thorough investigation of the hyperparameters across 11 datasets and we identify certain aspects that can be exploited. Analyzing the effects of certain hyperparameter settings on the empirical performance, as measured by cross validation accuracy, and generalization ability, as measured by geometric difference described above, brings us one step closer to understanding the potential of quantum kernel methods on classical datasets.
\end{abstract}


\section{Introduction}
Assessing the performance of quantum models on current quantum computers in tackling machine learning tasks on classical datasets is an enduring ambition of the research community. Many strides have been made addressing this subject from theory-driven approaches that showed encouraging results for \textit{specific models}, e.g., investigating the difference in capacity between classical and quantum models \cite{Abbas_2021} and showcasing a model that allows to sample from a circuit that is hard to simulate classically \cite{Coyle_2020}, to more empirical-driven projects that showed promising results for \textit{specific models} on \textit{specific use-cases}, e.g., medical classification with hybrid quantum-classical convolutional neural network \cite{monnet2023pooling} and generation of high-resolution MNIST datapoints with hybrid generative neural networks \cite{rudolph2022generation}. However, making an overall claim of quantum advantage for a \textit{broader class of models} for \textit{a broader set of use-cases} remains challenging, especially due to the notorious insufficiency of theoretical understanding of certain sub-fields of machine learning, such as deep learning (DL) \cite{Lin_2017}. Nevertheless, in the recent years, a body of literature emerged linking DL models to kernel methods (KMs) \cite{daniely2017deeper, jacot2020neural, lee2018deep} followed by literature making similar conclusion for the quantum domain \cite{rad2023deep, incudini2022quantum, schuld2021supervised}. This is great news due to favorable qualities of KMs, such as exhibiting convex loss landscapes, which allows to establish convergence, and hence trainability, guarantees \cite{schuld2021supervised}. It is a big advantage in comparison to DL that infamously suffer from non-convex loss landscapes and their quantum counterparts that suffers from barren plateaus \cite{McClean_2018}. Focusing on KMs therefore allows for a more theoretically sound argument, while at the same time, the insights gained for KMs can be applied to a broader range of models due to the aforementioned interrelationships.

\textit{Huang et al.}~\cite{Huang_2021} by arguing from a standpoint of generalization of kernels challenged a common reasoning for claiming quantum advantage that is often used in quantum machine learning (QML) literature. In their work, they showed that a classical intractability of the function, that is represented by a quantum model, is not sufficient to claim a quantum advantage. A classical learner can learn to approximate the quantum output if a sufficient amount of data is available. Following this argumentation, they presented a method that allows to probe specific quantum and classical kernels on a specific dataset to derive a possibility for quantum kernels to exhibit superior performance. The method consists of two steps: (1) a geometric difference (GD) measure is computed between kernel matrices, which allows to determine whether the kernels are different enough for potential quantum advantage to exist; (2) the model complexities of classical and quantum models are compared to determine if their discrepancy is sufficient (high classical and low quantum complexities are desirable).

In this work, we strive to comprehend the capacity of quantum kernels on classical datasets. It is our ambition to understand the possible gap between classical and quantum models and whether GD can be a helpful ally for this purpose. However, given a rather theoretical nature of this metric, we need to first deepen our grasp of the empirical relevance of the GD tool. To this extent, we study how the choice of hyperparameters influences the GD and the empirical performance of the quantum KMs. We performed an experiment spanning multiple different classical datasets, a large hyperparameter search grid, and different kernel setups. Additionally, we artificially saturate the difference between classical and quantum learners for some datasets by creating artificial quantum label function following ideas in \cite{Huang_2021} to expand our understanding of GD metric. We empirically established which hyperparameters QKMs are most sensitive to and hence are the most crucial to optimize. Our results indicate that the GD depends on the individual hyperparameters as much as the accuracy, but the optimal values of them differ greatly between the metrics. These insights open new research revenues towards understanding which use-cases might be most well suited for QKMs.

\section{Background}\label{sec:background}
Previous work by \textit{Moussa et al.}~\cite{Moussa_2022} investigated the importance of hyperparameters of \textit{quantum neural networks}, in which the performance of different hyperparameter settings has been evaluated on multiple datasets. From their experimental results, it follows that the choice of data embedding, the depth of the quantum circuit and the learning rate of the optimizer are the most important hyperparameters. In this work, we draw our inspiration from their experimental design, but we adapt our search grid to hyperparameters that are meaningful in \textit{quantum kernel methods} setting and conduct additional analysis to determine versatility of GD as potential metric for hyperparameter tuning. This section provides background information on our approach.

\subsection{Kernel functions}
A \textit{kernel function} $k(\mathbf{x},\mathbf{x}'): \mathbb{R}^D \times \mathbb{R}^D \to \mathbb{R}$ is bivariate, real-valued and can be interpreted as a similarity measure between ${\mathbf{x}, \mathbf{x}'} \in \mathbb{R}^D$, which is symmetric $k(x,x') = k(x',x)$ and non-negative $k(x,x')\geq0$. For any finite set $\mathcal{X}_N \subset \mathbb{R}^D$, we can construct a \textit{Gram matrix} defined by

$$K_{\mathcal{X}_N} = 
 \begin{pmatrix}
  k(\mathbf{x}_1,\mathbf{x}_1) & \cdots & k(\mathbf{x}_1,\mathbf{x}_N) \\
   & \vdots  &  \\
  k(\mathbf{x}_N,\mathbf{x}_1) & \cdots & k(\mathbf{x}_N,\mathbf{x}_N)
 \end{pmatrix},$$
which is Hermitian and positive semi-definite (PSD). According to Mercer's theorem, a Gram matrix with this property allows to represent the kernel function as an inner product of feature vectors $k(\mathbf{x},\mathbf{x}') = \phi(\mathbf{x})^T \phi(\mathbf{x}')$, where $\phi: \mathbb{R}^D \to \mathcal{H}$ \cite{MLprob} is a \textit{feature map} that maps original input space into a higher dimensional feature Hilbert space. 

Depending on a choice of a feature map $\phi$, we can generate kernels with different properties. For example, for $\mathbf{x} \in \mathbb{R}^D$ we can consider the simplest (linear) map $\phi(\mathbf{x}) = \mathbf{x}$, and we get a linear kernel defined as
    $k(\mathbf{x},\mathbf{x}') = \mathbf{x}^T \mathbf{x}'$.
One of the most widely-used maps $\phi(x)$ generates a \textit{radial basis function (RBF)} kernel defined as
\begin{equation}\label{eq:kernel_RBF}
    k(\mathbf{x},\mathbf{x}') = \exp(-\gamma||\mathbf{x} - \mathbf{x}'||^2 ),
\end{equation}
where $\gamma = \sigma^{-2}$ and $\sigma$ is known as \textit{bandwidth}. This kernel has infinite dimensions, which becomes apparent if we consider the Taylor expansion of \cref{eq:kernel_RBF}. 

Kernels are employed by a variety of machine learning algorithms. One of the most prominent example of such algorithms is a binary linear classifier known as a \textit{Support Vector Machine (SVM)}. The main objective of an SVM is to find a separating hyperplane between two classes of data with the maximal separating margin. However, the data might not necessarily be linearly separable in the input space $\mathbb{R}^D$, in which case mapping data into a higher dimensional feature space $\mathcal{H}$ is of an advantage. Moreover, one can avoid expensive explicit computation in high or infinite dimensional space by utilizing what is known as a \textit{kernel trick}. In essence, one computes merely an inner product of two feature vectors without accessing high dimensional space.

\subsection{Hamiltonian evolution feature map}
For a quantum algorithm to operate on a classical data, the data has to be embedded into a quantum state $\phi: \mathbf{x}_i \to \ket{\mathbf{x}_i}$, which exists in a complex Hilbert space $\mathcal{H}$ that is $\mathbb{C}^{2^n}$ with $n$ being the number of qubits. The significance of the choice of embedding operation for the performance of quantum models has been highlighted in the literature \cite{Abbas_2021, Schuld_2021}. In this work, we concentrate on the Hamiltonian evolution feature map as it has been proven to be powerful in QKM setting \cite{canatar2023bandwidth, Huang_2021, Shaydulin_2022}. This feature map has been developed for many-body problems, and it requires $n +1$ qubits to embed $n$ features and is defined as follows
\begin{equation}\label{eq:embedding}
    \ket{\mathbf{x}_i} = \bigg(\prod_{j=1}^n \exp \Big(-i \frac{t}{T}x_{ij} H_j^{XYZ} \Big)\bigg)^T \bigotimes_{j=1}^{n+1} \ket{\psi_j},
\end{equation}
where $H_j^{XYZ} = (X_j X_{j+1} + Y_j Y_{j+1} + Z_j Z_{j+1})$ with $X_j, Y_j, Z_j$ being Pauli operators acting on $j$-th qubit, $\ket{\psi_j}$ is a Haar-random state, and $t$ and $T$ are hyperparameters that symbolizes total evolution time and number of Trotter steps, respectively.
The significance of hyperparameter $t$ for QKMs has been highlighted by \textit{Shaydulin et al.}~\cite{Shaydulin_2022}. Our results support and extend their work.

\subsection{Quantum kernels}\label{sec:quantum_kernels}
An inner product of two quantum states defines a \textit{quantum fidelity kernel} as
\begin{equation}\label{eq:fidelity_kernel}
    k(x, x') = |\braket{\phi(x)}{\phi(x')}|^2.
\end{equation}
Quantum kernels insert data into exponentially high dimensional Hilbert feature space, which is intractable for classical computers. The high dimensionality of this space has to be treated with caution as it requires $N \geq \Omega(2^n)$ of training data to approximate a certain function well, where $n$ is the amount of logical qubits \cite{Huang_2021, kubler2021inductive, nakaji2022deterministic, thanasilp2022exponential}. This is due to a problem known as exponential concentration \cite{thanasilp2022exponential}. The same logic holds for a quantum fidelity kernel of mixed quantum states $k(x, x') = Tr(\rho(x) \rho(x'))$ \cite{nakaji2022deterministic}. Reducing the dimensionality of the space by projecting it onto a smaller subspace, thereby creating so-called \textit{projected quantum kernels}, increases the generalization capacity of these methods. To achieve that, we can utilize reduced density matrices (RDMs) $\rho^{(q)}(x) = Tr_{\neq q} (\Phi(\rho(x)))$, where $Tr_{\neq q}$ stands for a trace of a \textit{subsystem} that is created by excluding $q$ qubits and $\Phi$ is a trace-preserving map. Depending on the choice of the map $\Phi$, different proposals for projected kernels can be found in the literature \cite{Huang_2021, kubler2021inductive}. As generalized by \textit{Nakaji et al.}~\cite{nakaji2022deterministic}, these projected (RDM-based) kernels can be categorized into \textit{inner product based} defined as
\begin{equation}\label{eq:rdm_inner}
    k(x, x') = \sum^{N_K}_k \alpha_k Tr(\rho^{(q_k)}(x)\rho^{(q_k)}(x')),
\end{equation}
where $\alpha_k > 0$ are coefficients, and \textit{distance based} defined as
\begin{equation}\label{eq:rdm_distance}
    k(x, x') = \exp(-\gamma \sum^{N_K}_k \alpha_k ||\rho^{(q_k)}(x) - \rho^{(q_k)}(x')||^2_{F}),
\end{equation}
where the Frobenius norm is denoted by $||.||_F$, $\gamma>0$ is a hyperparameter, and $\alpha_k > 0$ are coefficients. The subsystem of size $K$ represented by $q_k$ is the \textit{k-th} of the $N_K$ possibilities.

\subsection{Geometric Difference (GD)}\label{sec:geometric_diff}
\textit{Huang et al.}~\cite{Huang_2021} challenged a common argument for claiming quantum advantage that is often used in QML literature. In their work, they showed that a classical intractability of the function $f(x)$, that is represented by a quantum model, is not sufficient to claim a quantum advantage. Under certain circumstances, a classical learner $h(x)$ can learn to approximate $f(x)$ given a sufficient number $N$ of independent samples from a data distribution $\mathcal{D}$. They showed that an expected error between $f(x)$ and $h(x)$ can be upper bounded as follows:
\begin{equation}~\label{eq:error_bound}
    \mathbb{E}_{\mathbf{x}\sim\mathcal{D}}|h(\mathbf{x}) - f(\mathbf{x})|\leq c\sqrt{\frac{s_K(N)}{N}},
\end{equation}
where $s_K(N)$ is the model complexity of the trained function $h(x)$ and $c>0$ is a constant. For that reason, with $N \propto s_K(N) / \epsilon^2$ datapoints a classical model can learn to predict the quantum function $f(x)$ up to an error $\epsilon$. Thus, if the best classical learner $h(x)$ has a low model complexity $s_K(N)$, it will not require too much data $N$ to learn the outputs of the quantum model $f(x)$. From here, it follows, that for a quantum advantage to exist it is required that there is the largest possible separation between complexities of quantum and classical models. \textit{Huang et al.}~\cite{Huang_2021} considered kernel models in their work and defined an asymmetric GD-based test between kernel matrices that captures this separation on a specific dataset. This test is defined as follows:
\begin{equation}~\label{eq:gscore}
    g_{CQ} = \sqrt{||\sqrt{K_Q}(K_C)^{-1}\sqrt{K_Q}||_{\infty}},
\end{equation}
where $K_C$ and $K_Q$ are classical and quantum kernel matrices with $Tr(K_Q)=Tr(K_C)=N$, respectively. The \textit{distance based} kernel meets this requirement, while the \textit{inner product based} kernel requires rescaling according to $\Tilde{K}_{Q} = N\cdot K_{Q}/Tr(K_{Q})$ in the case of projected kernels, due to the fact that the diagonal of this kernel's matrix contains purity related values (see \cref{app:inner_norm}). Importantly, this test is reliant on the GD between embedded datapoints, without considering the labeling function.
The GD now links the classical model complexity $s_{K_C}(N)$ and the quantum model complexity $s_{K_Q}(N)$ according to
\begin{equation}~\label{eq:gscore_ineq}
     c\sqrt{\frac{s_{K_C}(N)}{N}}\leq c g_{CQ}\sqrt{\frac{s_{K_Q}(N)}{N}},
\end{equation}
which is also a connection of the prediction error bounds as given in \cref{eq:error_bound}.

\section{Experimental setup}\label{sec:experiment}
In this work we investigate how different parameter settings affect the empirical performance of quantum kernels, as well as their generalization ability, as determined by the GD. We have designed an experiment to evaluate a variety of different hyperparameter settings. 
In this section, we describe the parameters of the search grid used in our experiment. We varied the types of models and their hyperparameters and tested their performance on various datasets.

\subsection{Models}
\label{sec:models}
In this work, we combine classical SVMs (see \cref{sec:background}) from \texttt{sklearn} library with quantum kernels implemented with \texttt{PennyLane} framework on simulators. To obtain the kernel, we first embed classical datapoints with the Hamiltonian evolution feature map as in \cref{eq:embedding} and the full DM of the resulting quantum state is saved. These DMs are used to compute RDM kernels described in \cref{eq:rdm_inner} (with additional rescaling to $Tr(K_Q)=N$), \cref{eq:rdm_distance} and \cref{eq:rdm_inner_norm}. From here forth, we will refer to these kernels as \textit{inner} and \textit{distance} kernels respectively. In all cases, we set $\alpha_k = 1/N_K$. In \cref{ap:alpha} we discuss the case $\alpha_k = 1$.

From these models we extract relevant hyperparameters. From SVMs we consider \textit{regularization} hyperparameter $C$. From the feature map we adopt the number of \textit{Trotterization steps} $T$ and the \textit{evolution time} $t$ (as in \cref{eq:embedding}). The random seed for the initial Haar-random state is set to different values, which however cannot be considered as a hyperparameter. From quantum kernels we take into account the \textit{size of the RDM subsystem} $K$ (as in \cref{eq:rdm_inner} and \cref{eq:rdm_distance}) and, in case of the \textit{distance} kernel, we also consider \textit{bandwidth} $\gamma$. All hyperparameters of the search grid and their corresponding value ranges are summarized in \cref{tab:hyperparameters}.

For a classical baseline we train SVM with classical kernels listed in \cref{sec:background}: from basic kernels (linear and polynomial) to more complex ones (RBF and Laplacian) and the (not necessarily PSD) sigmoid kernel. For this comparison the $C$ and $\gamma$ parameters are explored for the same range as in the quantum case with addition of more standard classical $\gamma$ choices, such as $1 / D$ and $1 / (D \cdot \sigma(\mathcal{X}^D))$ with $D$ being the dimensionality of the feature space and $\sigma(\mathcal{X}^D))$ variance of the dataset.
 
\begin{table}
\centering
\begin{tabular}{lclcc} 
\toprule
\textbf{Hyperparameter} & \textbf{Notation} & \textbf{Range} & \textbf{Number} & \textbf{Spacing}\\
\midrule
Evolution time & $t$ & $[2^{-6}, 2^6]$ & 13 & log\\

Trotterization steps & $T$ & $[1,3^4]$ & 5 & log\\

Bandwidth & $\gamma$ & $[10^{-3}, 10^{3}]$ & 13 & log\\

RDM size & $K$ & $[1, D + 1]$ & $D$ & linear \\

Regularization & $C$ & $[10^{-1}, 10^5]$ & 13 & log\\
\bottomrule
\end{tabular}
\caption{List of hyperparameters for embedding (Trotter steps $T$ and evolution time $t$), kernel function (subsystem size $K$ and bandwidth $\gamma$) and SVM (regularization $C$) and their corresponding values range, number of samples and spacing that were used in the search grid. $D$ is the number of features.}
\label{tab:hyperparameters}
\end{table}


\subsection{Metrics}
To rate the importance of the hyperparameters, we need to correlate them with the model's performance. In this paper, we consider performance metrics, such as \textit{accuracy on test dataset} and \textit{mean accuracy of 5-fold cross validation on training dataset}, as well as the generalization metric \textit{geometric difference (GD)} introduced in \cref{sec:geometric_diff}. GD measures the difference between two kernel matrices, so in our studies we consider the GD from quantum to all classical matrices.

\subsection{Datasets} \label{sec:datasets}
To train and test our models, we selected the 11 different datasets listed in \cref{tab:data} that have been previously used in the QML literature \cite{Huang_2021, moradi2022clinical, Moussa_2022, Shaydulin_2022}.
Due to simulation time restrictions, we compress each dataset's feature space to 5 dimensions, which translates to 6 qubits according to \cref{eq:embedding}, and 200 datapoints (or less features/datapoints if fewer were available). In this study, we restrict to binary classification tasks, so for multi-class dataset 8 (Fashion-MNIST) only two classes are selected (T-shirt/Top and Dress) and the rest are discarded. Datapoints with incomplete or missing features are removed as well as duplicate datapoints. In addition, the datasets were balanced. Features with a variance of less than 0.001 were eliminated. All the features were normalized (mean with 0 and variance with 1). To reduce the dimensionality of the features, all features were sorted according to their analysis of variance (ANOVA) F-value, and the top 5 were selected. Different preprocessing routines are discussed in \cref{ap:preprocessing}.

\begin{table}
\centering
\begin{tabular}{clc} 
\toprule
\textbf{ID} & \textbf{Name} & \textbf{Source} \\
\midrule
\multirow{1}{*}{Dataset 1} & Breast Cancer Wisconsin (Original) & \multirow{1}{*}{\cite{bischl2021openml}}/\cite{ucimlrepo} \\ 

\multirow{1}{*}{Dataset 2} & Pima Indians Diabetes Database & \multirow{1}{*}{\cite{bischl2021openml}} \\ 

\multirow{1}{*}{Dataset 3} &Banknote Authentication & \multirow{1}{*}{\cite{bischl2021openml}} \\ 

\multirow{1}{*}{Dataset 4} & Blood Transfusion Service Center Data Set & \multirow{1}{*}{\cite{bischl2021openml}} \\ 

\multirow{1}{*}{Dataset 5} &Indian Liver Patient Dataset & \multirow{1}{*}{\cite{bischl2021openml}} \\ 

\multirow{1}{*}{Dataset 6} &Phoneme data set & \multirow{1}{*}{\cite{bischl2021openml}} \\ 

\multirow{1}{*}{Dataset 7} &Wilt Data Set & \multirow{1}{*}{\cite{bischl2021openml}} \\ 

\multirow{1}{*}{Dataset 8} &Fashion MNIST & \multirow{1}{*}{\cite{bischl2021openml}}/\cite{TF_quantum} \\ 

\multirow{1}{*}{Dataset 9} &Breast Cancer Wisconsin (Diagnostic) & \cite{ucimlrepo}\\ 

\multirow{1}{*}{Dataset 10} &Pediatric Bone Marrow Transplant & \cite{ucimlrepo}\\ 

\multirow{1}{*}{Dataset 11} &Heart failure clinical records & \cite{ucimlrepo}\\ 
\bottomrule
\end{tabular}
\caption{A comprehensive list of all datasets and their sources that were used in the experiment. If there are 2 possible sources for a dataset and both have been used, both are indicated.}
\label{tab:data}
\end{table}

\subsection{Artificial labels}\label{sec:artificial labels}
For a more in-depth study of the GD, we additionally employ a method for generating artificial labels for classical datasets that are more favorable for quantum learners as proposed in \cite{Huang_2021}. This is done by first augmenting \cref{eq:gscore} to a matrix of form $\sqrt{K_Q}(K_C + \lambda Id)^{-1}\sqrt{K_Q}$ with the regularization term $\lambda = 1.1$. We select eigenvector $\mathbf{v}$ that corresponds to the largest absolute eigenvalue, and the intermediate new labels are calculated as $y = \sqrt{K_Q}\mathbf{v}$. These labels are binarized by setting them to 1 if they are greater than the median of all $y$ and to 0 otherwise. The learning task with these new labels should now saturate \cref{eq:gscore_ineq}. However, the kernel matrices in this scenario are formed using the complete dataset, resulting in slight deviations from this saturation upon separation into training and test sets. To improve generalization, 5\% of the labels are randomly set to 0. We use different Haar random initial states for relabeling and learning. Most of the analysis is performed on initial datasets. The relabeled ones are discussed in \cref{sec:rela}.

\subsection{Analysis}\label{sec:analysis}
After all results from the experiments are agglomerated, the data is prepared for further analysis. The choice of the kernel function is treated as a hyperparameter. The two kernel functions are encoded as the binary variable \textit{basis}. We also exclude hyperparameter $C$ (as it has no influence on the GD) from the main analysis and discuss it separately in \cref{sec:c}. For datasets 1 and 8, we removed the highest 3\% of each GD due to the presence of significant outliers. These outliers may occur when the classical kernel matrix has nearly singular values (\cref{eq:gscore}) and the GD is therefore exceedingly large.

We start the analysis by investigating the importance of each hyperparameter. 
Based on the insights from our prior experiments, we chose the Gradient Boosted Decision Trees (GBDT) from \texttt{sklearn}, 
which is fitted to the experimental data. The features correspond to the hyperparameters and the labels to the accuracies of the earlier classification tasks. Such a GBDT consists of hundreds of decision trees. Depending on how often a certain feature (hyperparameter) is used for a split point, a feature importance can be derived. More precisely, from this model an impurity-based feature importance, also known as the Gini importance, can be extracted.

Additionally, we analyze marginal distributions of each hyperparameter, which is an average over all other hyperparameters to capture the effects of this particular one on the accuracy and the GD to the classical RBF kernel. Apart from the mean, we also track standard deviation as it provides the necessary information regarding reliability of the marginal distribution. A large standard deviation indicates a small importance of that particular hyperparameter on the metric and vice versa. The impact of each hyperparameter on performance and GD is evaluated using the GBDT and its margin plots.


\section{Results}\label{sec:results}
\begin{figure}
  \centering
  \includegraphics[width=\linewidth]{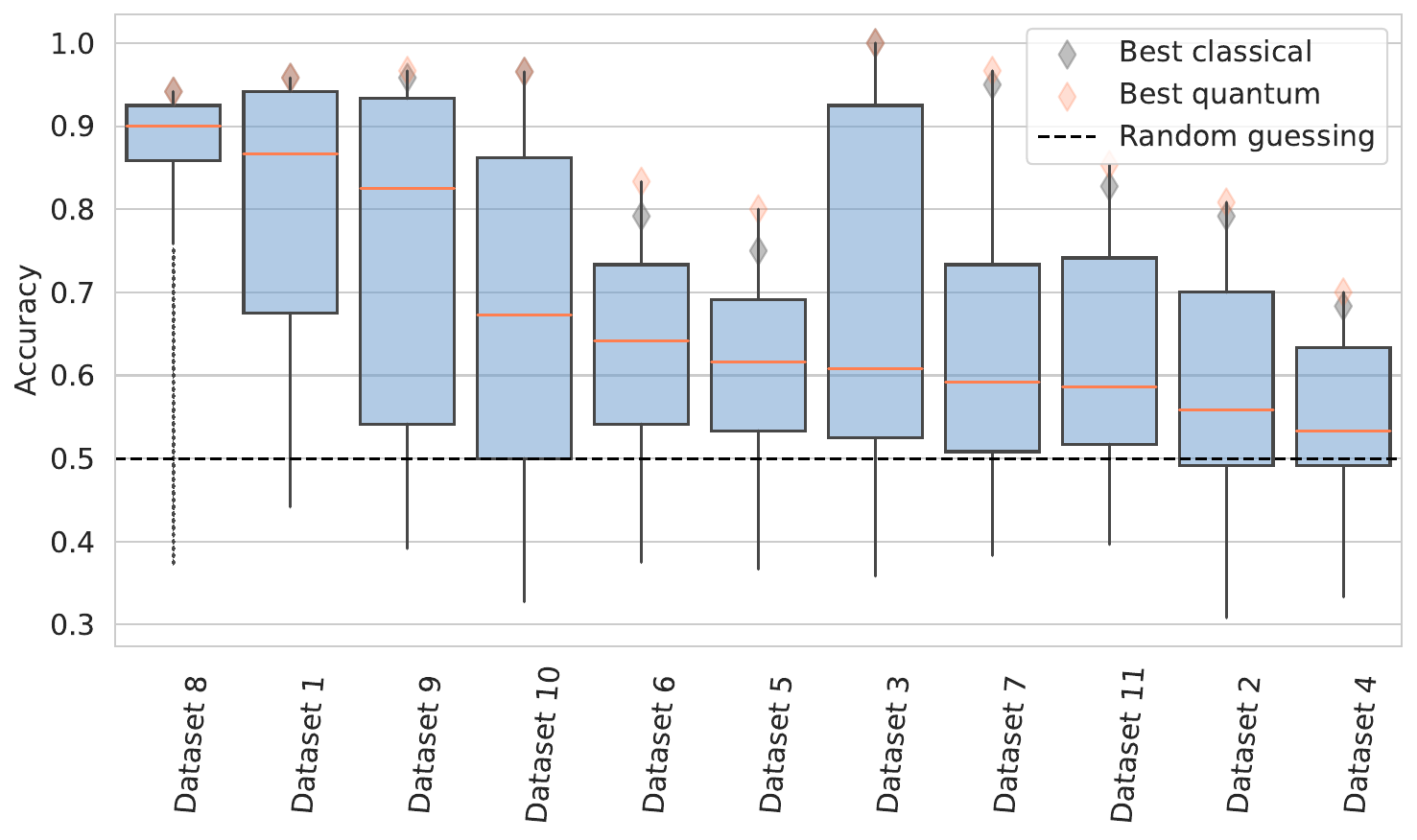}
  \caption{Distribution of accuracies achieved by an SVM with a quantum kernel on the test set for all hyperparameters setting from the search grid (\cref{sec:models}) for each dataset (\cref{tab:data}). The distributions are visualized as boxplots. Additionally, the diamonds indicate the best accuracies achieved by classical and quantum models. The dashed line indicates the accuracy that would be achieved by random guessing.}
  \label{fig:acc_over_dataset}
\end{figure}

\cref{fig:acc_over_dataset} visualizes the range of accuracies across different search grid settings as well as best achieved performances of classical and quantum learners (for more details on best learners see \cref{ap:performances_by_models}). Since the quantum models achieve comparable results to the classical counterparts, we can assume that the explored hyperparameter ranges allowed for a high enough expressivity. Nevertheless, the best classical model can barely be surpassed. The selection of datasets and their characteristics range from those which are easy to learn, to those which are difficult to learn. There is also a varying spread of the accuracies on one specific dataset across the different parameter settings. 
\subsection{Hyperparameter importance}\label{sec:importance}
\begin{figure}
  \centering
    \includegraphics[width=\linewidth]{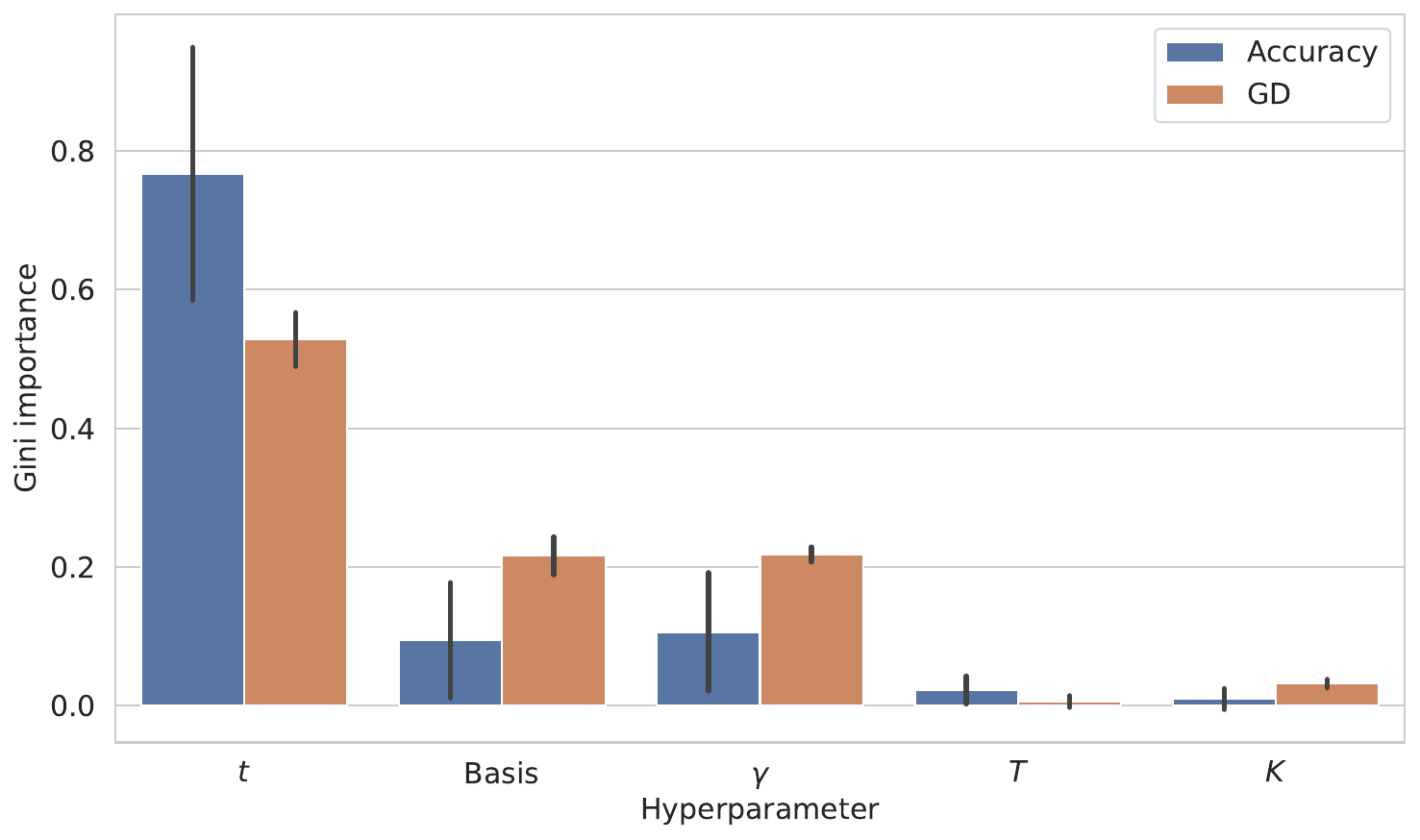}
    \caption{Mean Gini importance for each hyperparameter listed in \cref{tab:hyperparameters}, across all datasets listed \cref{tab:data} and for both accuracy and GD to the classical RBF kernel. The height of each bar represent mean value while error bars indicate standard deviation. The importance indicates how much influence the variation of this hyperparameter value has on one of the two metrics. These importances were extracted from GBDT model fitted on the results from the search grid (\cref{sec:analysis}).}
    \label{fig:importance_barchart}
\end{figure}

We analyze whether accuracy and the GD metrics assign the same importance to the hyperparameters. For that, we extract Gini importance from the GBDT model fitted to the data extracted from the search grid experiment as described in \cref{sec:analysis}. 
\cref{fig:importance_barchart} displays the average importance found for the different hyperparameters across all the datasets for performance as well as for the GD to the classical RBF kernel. See \cref{fig:Gini_all_metrics} for a breakdown of hyperparameter importance by all metrics.

Contrary to our initial expectations, both the performance and GD exhibit minimal sensitivity to variations in the value of $K$. Also, the number of Trotterization steps $T$ has no significant effect on any of the metrics studied. These results are consistent with \cite{Shaydulin_2022}, where $t$ and $T$ hyperparameters were investigated as well. This analysis also found that the random seed for the Haar initial state had the lowest importance for every metric, but as it cannot be interpreted as a hyperparameter, it will not be discussed further.

Judging from the results in \cref{fig:importance_barchart}, the hyperparameter $t$ is assigned the most importance by both the accuracy and the GD metrics. The choice of the \textit{basis} and $\gamma$ has a strong effect on these metrics as well, as both assigned these two parameters a similar importance that ranks them together at the second place. When examining the standard deviations, it is apparent that the importance of each hyperparameter for the GD is more consistent across the datasets than the importance for the accuracy.
\subsection{Metric dependence on individual hyperparameters}\label{sec:indi_params}
In this section, the effects of individual hyperparameters are discussed. For the most part, we base our assessments on the marginals of the hyperparameters (see \cref{sec:analysis} for further details). The hyperparameters are analyzed in the order of their importance. As expected, the cross validation (CV) aligns closely with the test accuracy in all marginals and is therefore well suited for tuning all hyperparameters.

\subsubsection{Evolution time ($t$)}\label{sec:time}
\begin{figure}
  \centering
  \includegraphics[width=\linewidth]{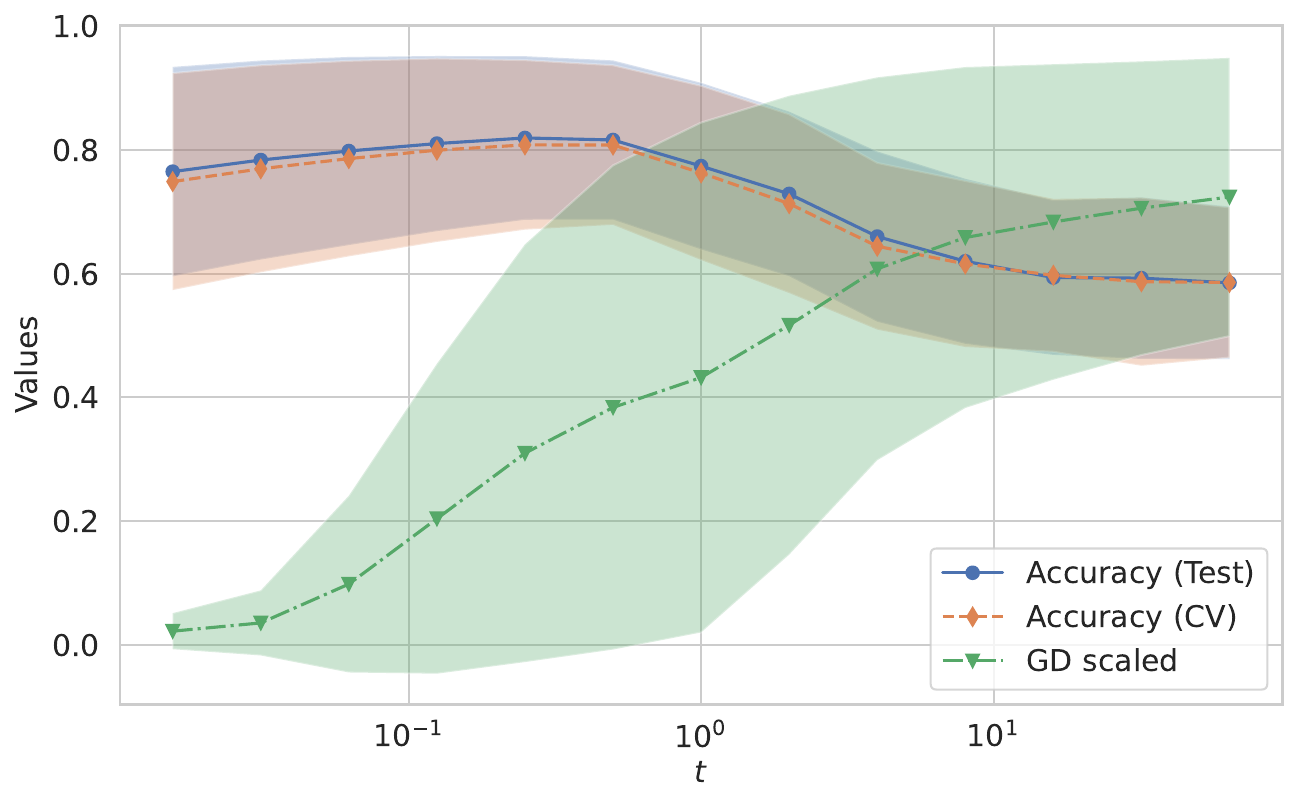}
  \caption{Dependence of the GD, the test accuracy and the cross validation accuracy on $t$ (evolution time). The shadows illustrate the standard deviations when averaging over the other parameters and over all datasets. The GD was scaled to the range $[0, 1]$ by dividing through the maximum. }
  \label{fig:time_marginal}
\end{figure}

\begin{figure*}
    \centering
    \begin{subfigure}[b]{0.28\textwidth}
        \includegraphics[height=10cm]{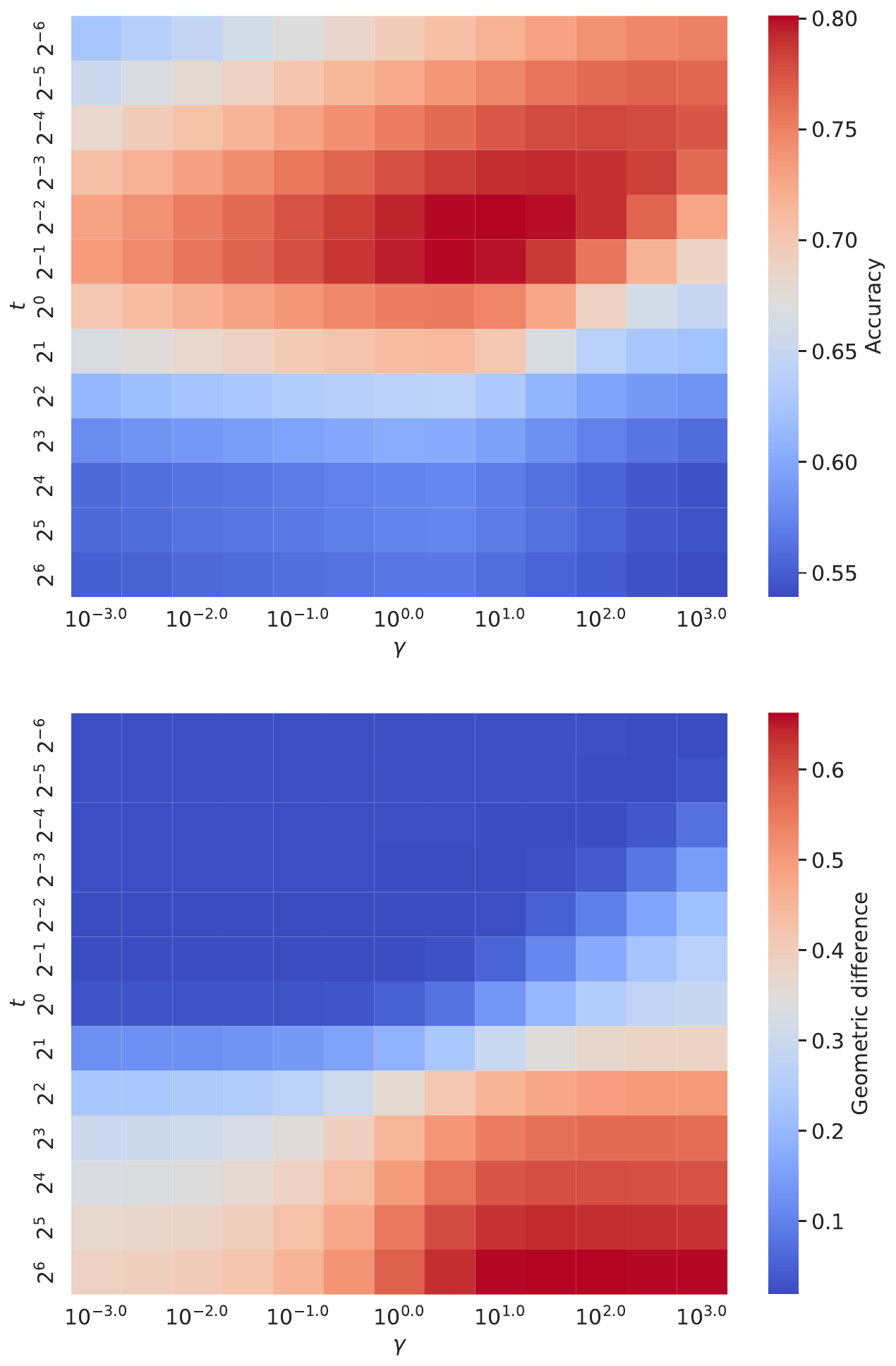}
        \caption{}
        \label{fig:acc_and_gs_heatmap}
    \end{subfigure}
    \hfill
    \begin{subfigure}[b]{0.64\textwidth}
        \includegraphics[height=10cm]{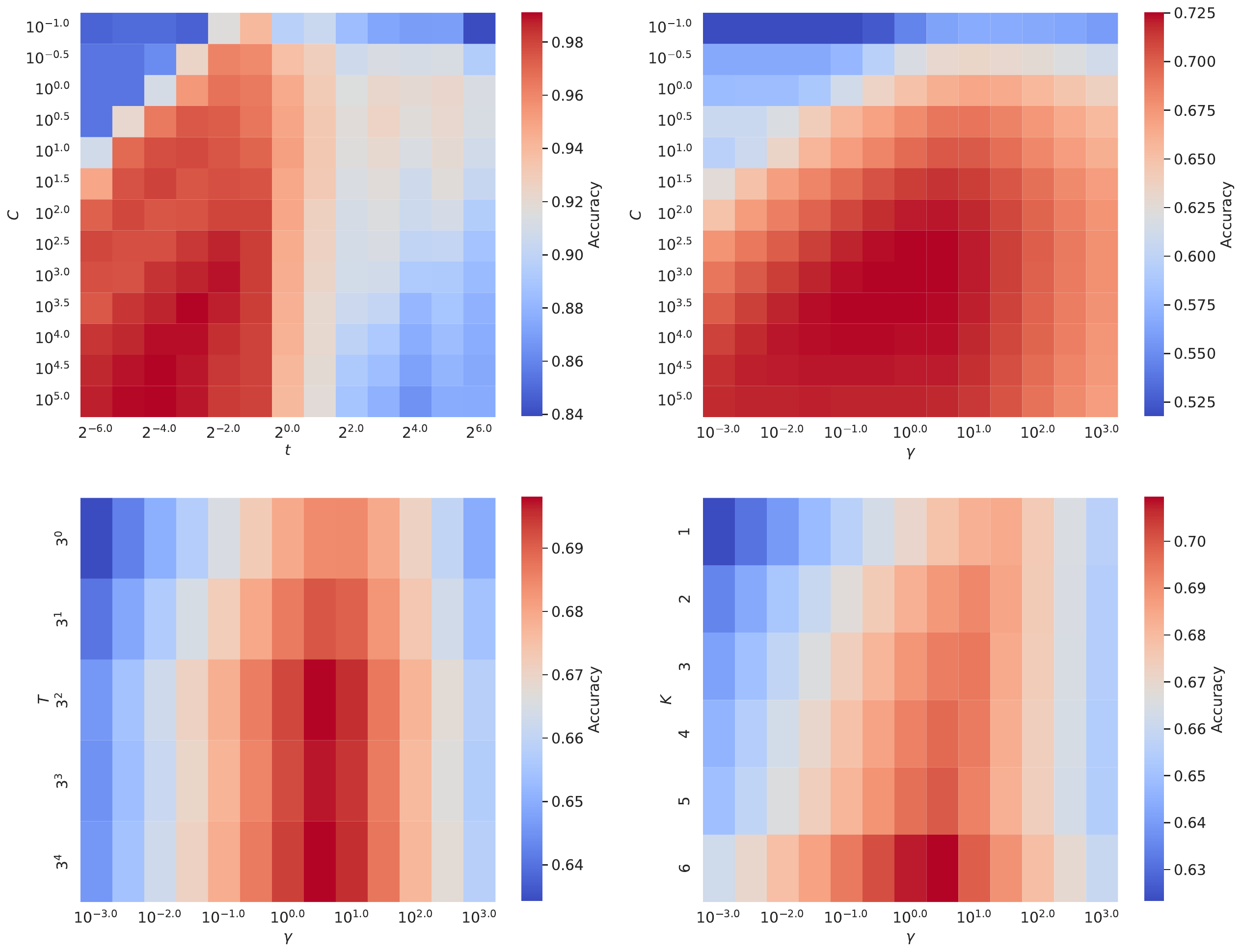}
        \caption{}
        \label{fig:heat_gamma}
    \end{subfigure}
    \caption{\textbf{a} Dependence of the accuracy and the GD on $\gamma$ (bandwidth) and $t$ (evolution time) combined. The graphic displays the mean over all datasets and other parameters.
    \textbf{b} Different marginals of the accuracy depending on 2 hyperparameters. $\gamma$ in combination with $C$ (regularization strength), $T$ (number of Trotterization steps) and $K$ (RDM size) as well as $t$ (evolution time) in combination with $C$ (regularization strength). 
    }
    \label{fig:heatmaps}
\end{figure*}

\cref{fig:time_marginal} visualizes the marginals of the $t$ parameter for the accuracy and the GD. The standard deviations are illustrated as shadows and indicate the importance of $t$ on each of those metrics. If a standard deviation is small when averaging over all other parameters and datasets, this indicates that the metric does not depend on these other parameters as strongly as on $t$. This graphic highlights that the $t$ values for which a good performance was achieved correspond to a rather small GD, and as the predictions get worse, the potential for quantum advantage increases. This behavior is consistent and sometimes even more significant when looking at individual datasets (\cref{fig:app_acc_time} and \cref{fig:app_GD_time}) or 2-marginals like \cref{fig:acc_and_gs_heatmap}. 
Furthermore, of all hyperparameters, $t$ causes the most differences between the choices of \textit{basis}. The \textit{distance} kernel seems to be more resilient for smaller values of $t$, demonstrated in \cref{fig:app_acc_time_dist_inner}. As $t$ increases, the GD begins to rise earlier when the \textit{distance} kernel is used.

\subsubsection{(\textit{Basis}) of the kernel function}\label{sec:basis}
\begin{figure}
  \centering
  \includegraphics[width=\linewidth]{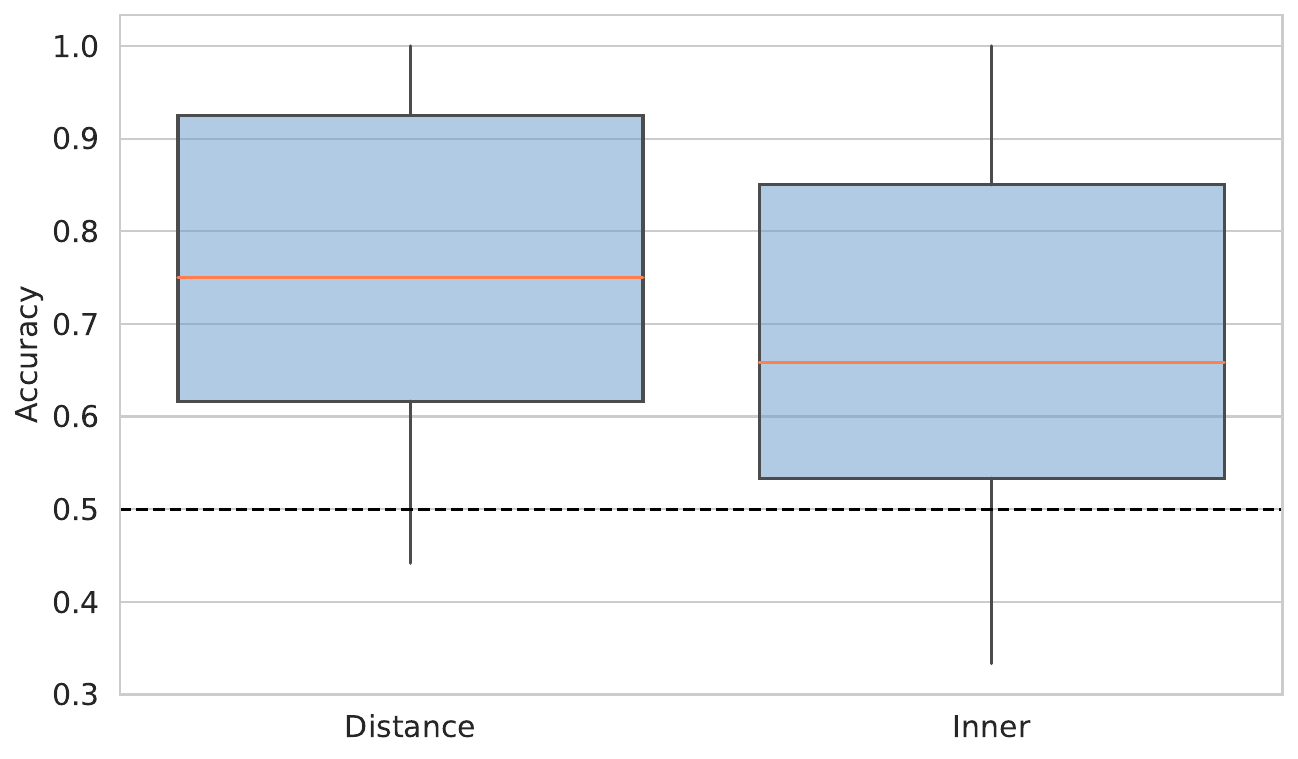}
  \caption{Dependence of the test accuracy on the choice of \textit{basis} for the kernel function. For this comparison $\gamma$ is optimized and not averaged. The specific form of a \textit{distance}- or \textit{inner} based kernel function is given in \cref{sec:quantum_kernels}. The dashed line indicates the accuracy that would be achieved by random guessing.}
  \label{fig:kernel_marginal}
\end{figure}

In order to have a fair comparison between the \textit{distance} kernel and the \textit{inner} kernel the $\gamma$ parameter is not averaged over, rather the maximal value over this parameter is chosen in all of \cref{sec:results}, where the choice between the two is considered a parameter too (\cref{sec:analysis}). This is further motivated by the fact that $\gamma$ along with $C$ are the computationally cheapest hyperparameters to optimize, since the kernel can be reevaluated without having to recompute the distances between DMs. 
\cref{fig:kernel_marginal} shows the accuracies achieved for the specific choices of kernel function \textit{basis} over all other parameter settings (except for $\gamma$) and datasets. If $\gamma$ is optimized, then the \textit{distance basis} outperforms the other two. In \cref{fig:app_kernel_marginal_gamma_avg}, we observe a different situation when $\gamma$ is averaged like all the other parameters, but this is mostly due to the broad range (and thus many suboptimal values) we explored.

\subsubsection{Gamma ($\gamma$)}\label{sec:gamma}
\begin{figure}
  \centering
  \includegraphics[width=\linewidth]{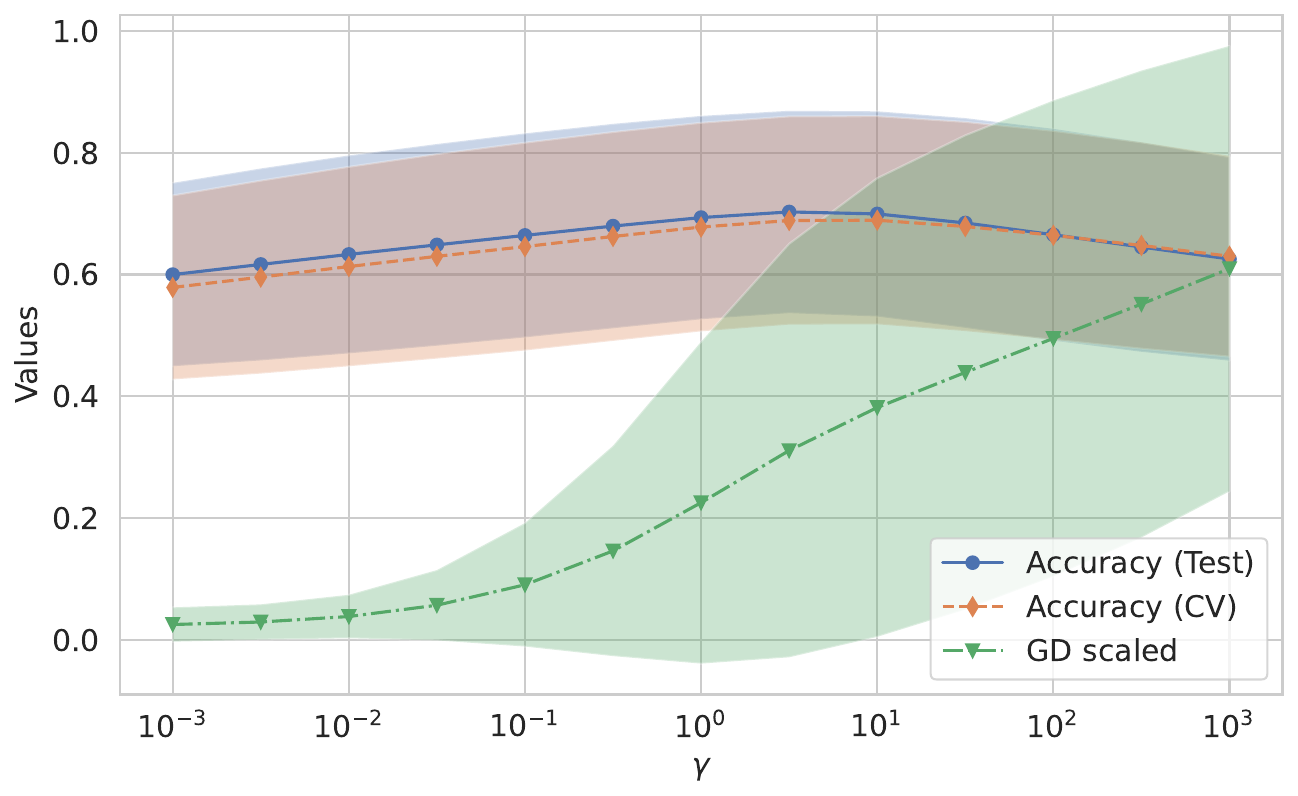}
  \caption{Dependence of the GD, the test accuracy and the cross validation accuracy on $\gamma$ (bandwidth). The shadows illustrate the standard deviations when averaging over all datasets and over the other parameters except for \textit{basis}, which is restricted to just the \textit{distance} kernel. The GD was scaled to the
range $[0, 1]$ by dividing through the maximum.}
  \label{fig:acc_over_gamma}
\end{figure}

The second hyperparameter controlling the bandwidth is gamma is $\gamma$. This parameter is only relevant for the \textit{distance} kernel (\cref{eq:rdm_distance}), which is why \cref{fig:acc_over_gamma}, \cref{fig:app_acc_gamma} and \cref{fig:app_GD_gamma} are restricted to just the \textit{distance} kernel. In contrast to the behavior seen in \cref{sec:time} the GD is here not necessarily running opposite to the accuracy, which can be seen from \cref{fig:acc_over_gamma} (or \cref{fig:app_acc_gamma} and \cref{fig:app_GD_gamma}). There is a general behavior that can be observed for all datasets, which is that all the GDs are increasing with $\gamma$. For most datasets the accuracy has a maximum in the range $0.1-10$ with varying curvature around that point across the different datasets. 
$\gamma$ is correlated with other parameters as seen in \cref{fig:acc_and_gs_heatmap} and \cref{fig:heat_gamma}. For example, the optimal value for $t$ shifts to smaller values with larger $\gamma$, which makes sense as the kernel values in general decrease when either of these parameter increases. 

\subsubsection{Size of the reduced density matrices ($K$)}\label{sec:rdm}
\begin{figure}
  \centering
  \includegraphics[height=5cm]{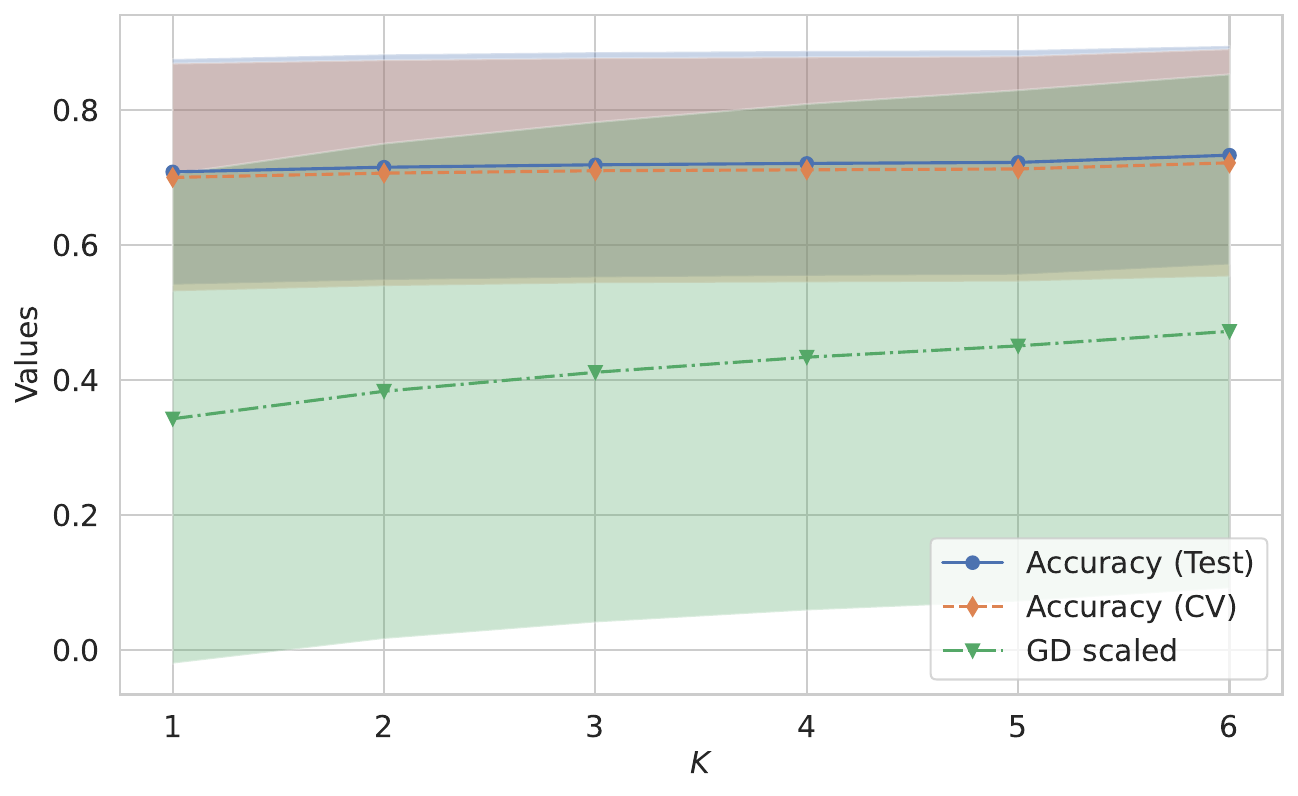}
  \caption{Dependence of the GD, the test accuracy and the cross validation accuracy on $K$ (RDM size). The shadows illustrate the standard deviations when averaging over the other parameters and over all datasets except for 3 and 4. The GD was scaled to the
range $[0, 1]$ by dividing through the maximum.}
  \label{fig:acc_over_rdm}
\end{figure}

Here, we are investigating RDMs that have the means to tackle the exponential concentration problem \cite{thanasilp2022exponential}. Because datasets 3 and 4 have only 4 features, they are excluded from all RDM-marginals in \cref{sec:results} as they would bias $K=6$. The trend in our results indicates a slight increase in accuracy with this hyperparameter, but minimal effect overall. While GD increases notably with $K$ as displayed in \cref{fig:heat_gamma} and \cref{fig:acc_over_rdm}. Both observations are consistent across all datasets (\cref{fig:app_acc_K} and \cref{fig:app_GD_K})). Furthermore, \cref{fig:importance_barchart} also indicates the difference in importance.

In \cref{ap:alpha}, we explore an alternative option for $\alpha_k$ in equations \cref{eq:rdm_inner} and \cref{eq:rdm_distance}, which primarily impacts the influence of $K$ on the metrics. Furthermore, artificial labels for various $K$ lead to different observations depending on $\alpha_k$.

\subsubsection{Number of Trotterization steps ($T$)}\label{sec:nTrotter}
\begin{figure}
  \centering
  \includegraphics[height=5cm]{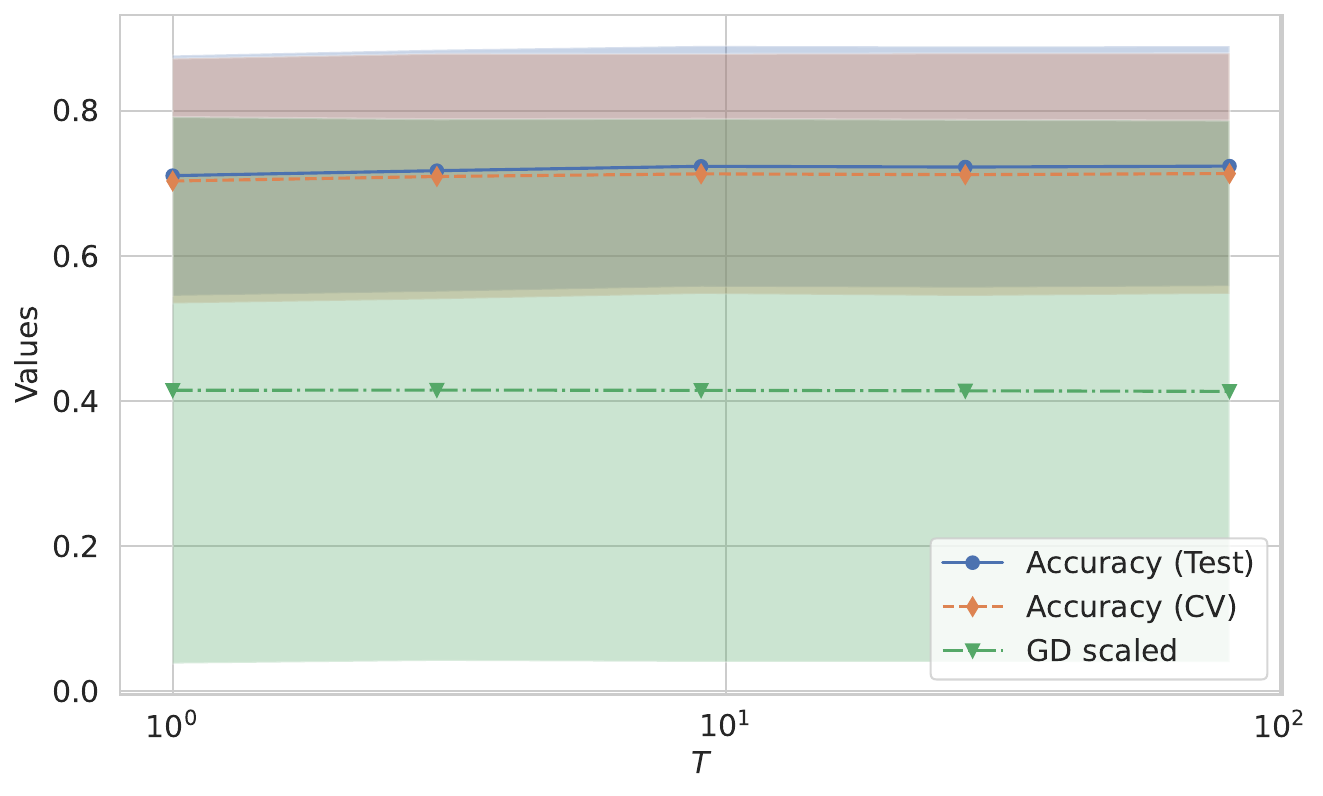}
  \caption{Dependence of the GD, the test accuracy and the cross validation accuracy on $T$ (number of Trotterization steps). The shadows illustrate the standard deviations when averaging over the other parameters and over all datasets. The GD was scaled to the
range [0, 1] by dividing through the maximum.}
  \label{fig:acc_over_n_Trotter}
\end{figure}

We found that varying $T$ has minimal impact on performance or GD for most datasets, which can be deduced from the large standard deviations in \cref{fig:acc_over_n_Trotter} and the almost absent variation of any metric with increasing $T$ (see also \cref{fig:app_acc_T} and \cref{fig:app_GD_T}). In \cref{fig:app_acc_T} only dataset 9 shows more accuracy with higher $T$.

\subsubsection{Regularization strength ($C$)}\label{sec:c}
\begin{figure}
  \centering
  \includegraphics[width=\linewidth]{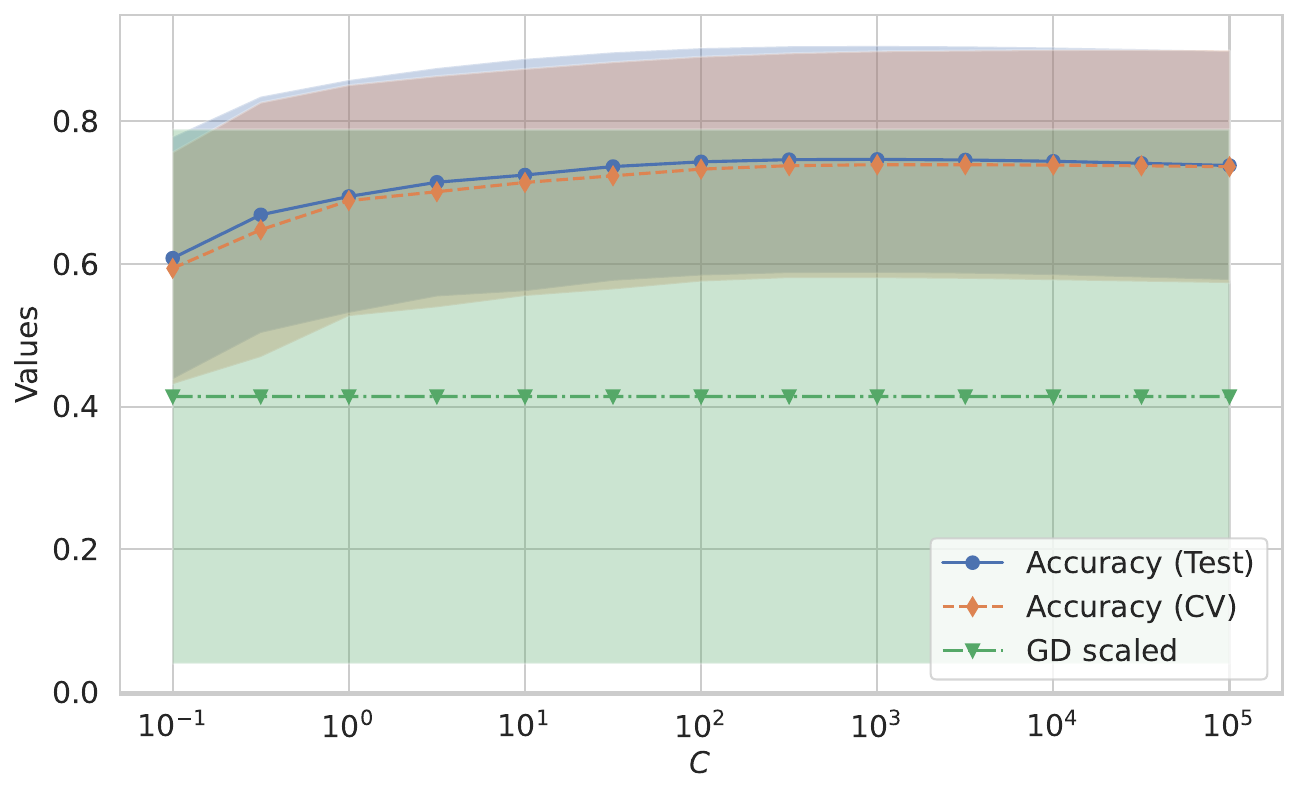}
  \caption{Dependence of the GD, the test accuracy and the cross validation accuracy on $C$ (regularization strength). The shadows illustrate the standard deviations when averaging over the other parameters and over all datasets. The GD was scaled to the range $[0, 1]$ by dividing through the maximum.}
  \label{fig:acc_over_c}
\end{figure}

Because $C$ does not affect the kernel, there is no dependence of the GD on it. The accuracy however has a notable dependence especially for smaller values (\cref{fig:acc_over_c}). The pattern is as expected for this parameter, which is that there is a strong increase in performance up to a global maximum. From there on the quality of the predictions slightly decreases with increasing regularization. There are large discrepancies between the datasets as to where that saturation happens and the slopes before and after (compare \cref{fig:app_acc_C}). 


\subsection{Relabeled datasets}\label{sec:rela}
With the method described in \cref{sec:artificial labels} we relabeled datasets for different quantum kernels. The motivation behind this is to detect differences between the original datasets and datasets that are more favorable for the quantum methods. Since the GD does not depend on the labels, this solely affects the accuracy. As the initial state has a negligible influence on the results (\cref{sec:importance}), we limit ourselves to a single random state in this section.

\begin{figure}
  \centering
  \includegraphics[width=\linewidth]{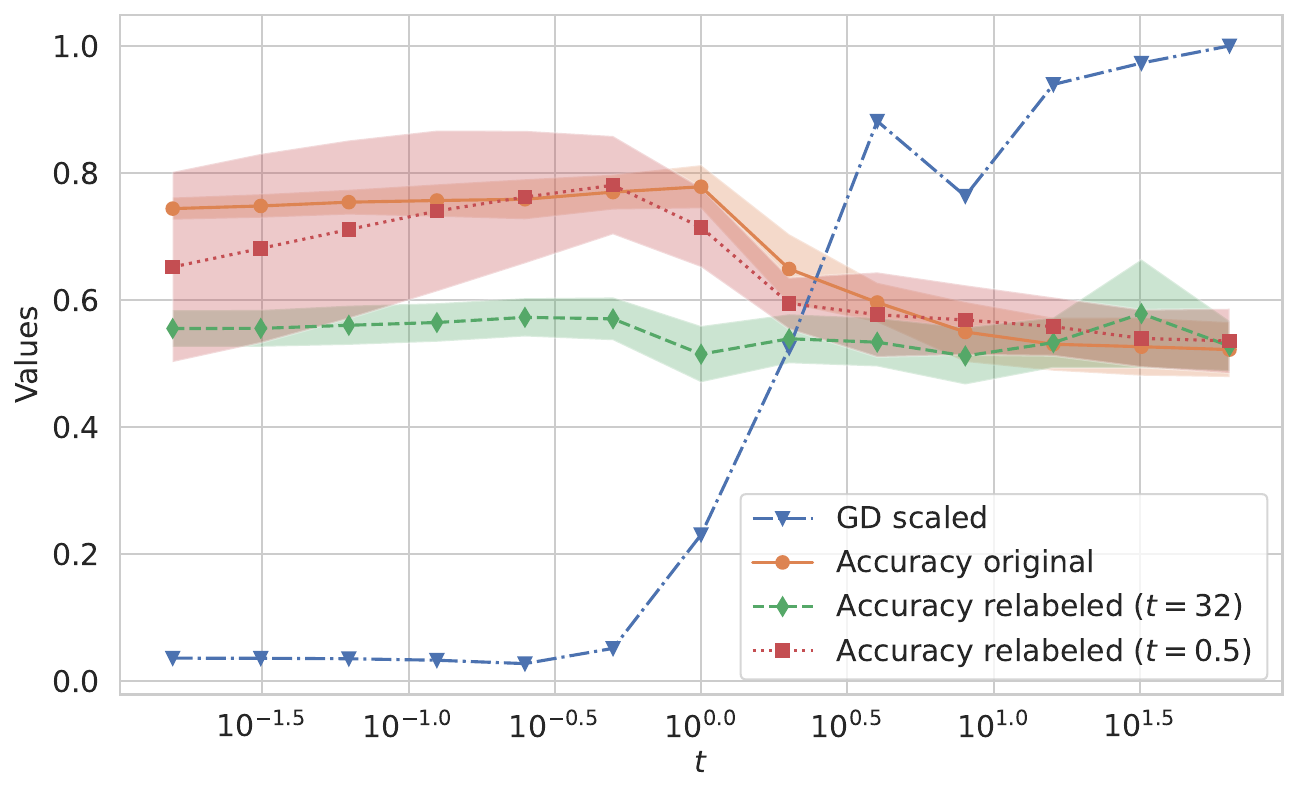}
  \caption{Dependence of the test accuracy on $t$ (evolution time) for the original dataset 6 and 2 relabeled versions for the settings: $basis = distance$, $T=9$, $K=6$, $\gamma = 1$ and either $t = 0.5$ or $t = 32$. The dependence of the GD on $t$ and for the specific choices of the other parameters is also shown. This metric is universal for the three versions, as the GD does not depend on the labels. Furthermore, it was scaled to the range $[0, 1]$ by dividing through the maximum. The shadows of the accuracies illustrate the standard deviations when averaging over the other parameters.}
  \label{fig:time_marginal_rela}
\end{figure}

We found that relabeling shifts the optimal hyperparameter values well to the setting for which the relabeling was performed. As an example, \cref{fig:time_marginal_rela} shows the marginal accuracies for the original dataset 6 compared to 2 relabeled versions for the settings: $basis = distance$, $T=9$, $K=6$, $\gamma = 1$ and either $t = 0.5$ or $t = 32$. Furthermore, the GD is plotted for exactly these choices of \textit{basis}, $T$, $K$, and $\gamma$. The marginal dependency on $t$ illustrates that the relabeling has shifted the optimal values for this hyperparameter to the settings for which the relabeling was performed. This effect was also observed for other parameters.

When the relabeling is done for $t=0.5$ the separation between the best classical and the best quantum accuracy is smaller compared to relabeling for $t=32$, which is displayed in \cref{tab:rela}. This is expected because the gradient GD increases with $t$ as shown in \cref{fig:time_marginal_rela}. Similarly, higher values of $K$ also result in larger GD (see \cref{sec:rdm}) in the case of $t=32$ this results in a larger separation. However, when $t=0.5$, the GD is too small to significantly improve the quantum model's ability to learn the artificial labels in comparison to the classical models. Small deviations in the separation can be explained by the process described in \cref{sec:artificial labels}. This process includes applying a regularization term ($\lambda Id$), utilizing a train-test-split, and incorporating additional randomness, which could explain any variance in the separation. Similar to \cref{tab:rela} we discuss \cref{tab:rela_alpha_1} in \cref{ap:alpha}, where we set $\alpha_k = 1$ instead of $\alpha_k = 1/N_K$ in \cref{eq:rdm_inner} and \cref{eq:rdm_distance}.

\begin{table}
\centering
\begin{tabular}{cccccc} 
\toprule
\textbf{K} & \textbf{t} & \textbf{Best classical} & \textbf{Best quantum} & \textbf{Separation} & \textbf{GD}\\
\midrule
1 & 32 & 0.63 & 0.64 & 0.02 & 1137.29\\
4 & 32 & 0.56 & 0.71 & 0.15 & 2059.87\\
6 & 32 & 0.65 & 0.81 & 0.16 & 2619.80\\
1 & 0.5 & 0.83 & 0.83 & 0.00 & 3.53\\
4 & 0.5 & 0.88 & 0.86 & -0.02 & 4.56\\
6 & 0.5 & 0.88 & 0.87 & -0.01 & 8.34\\
\bottomrule
\end{tabular}
\caption{List connecting the difference between the best classical and the best quantum accuracies and their separation with the GD. The columns $K$ and $t$ indicate the hyperparameter values for which dataset 6 was relabeled. The other parameters were set to $basis = distance$, $T=9$, $K=6$, $\gamma = 1$ and $\alpha_k = 1/N_K$ (\cref{eq:rdm_inner} and \cref{eq:rdm_distance})}
\label{tab:rela}
\end{table}
\begin{figure*}
    \centering
    \includegraphics[width=\textwidth]{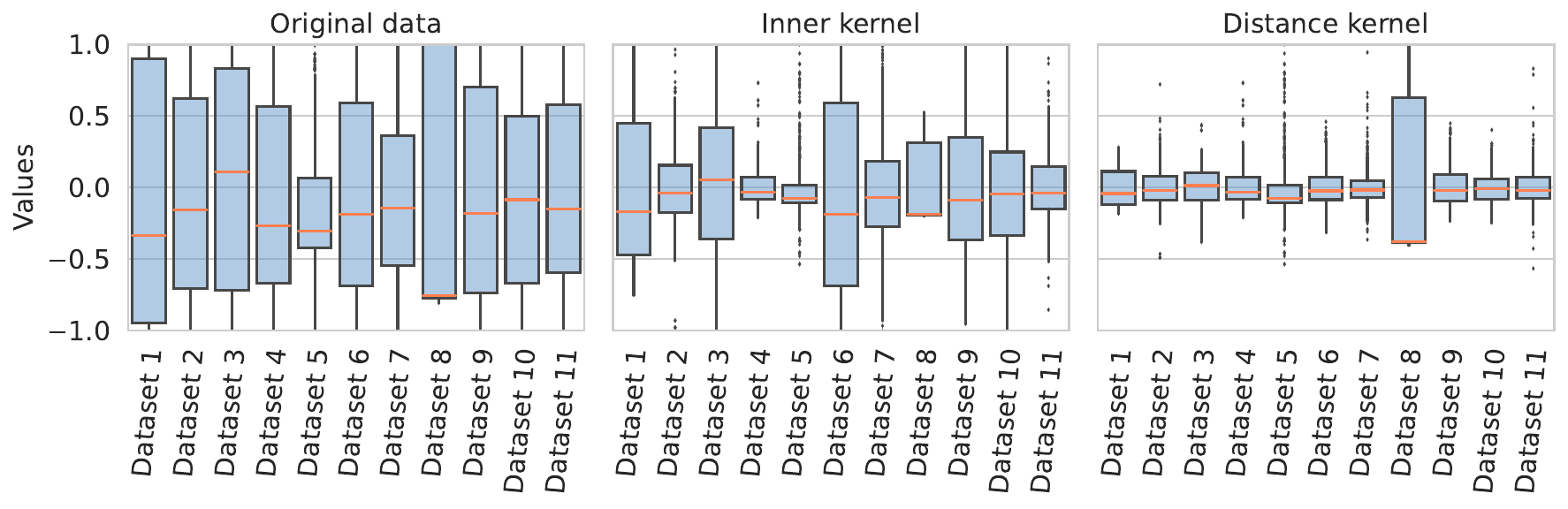}
    \caption{All datasets from \cref{tab:data} and their original feature scales (left). The features are scaled by the best on average $t$ values, which is the parameter that is responsible for the dataset scaling. We consider \textit{inner} (middle) and \textit{distance} (right) kernels separately, since the \textit{distance} kernel has the additional $\gamma$ hyperparameter that controls bandwidth too, which is closely related to the parameter scaling.}
    \label{fig:data_scaling}
\end{figure*}

\section{Pipeline for investigating new datasets}\label{sec:pipeline}
In this section, we agglomerate lessons-learned into a coherent pipeline for investigating new datasets with QKMs in combination with the Hamiltonian evolution feature map. These insights are specific to our experimental setup (6 qubits systems in a simulation environment, dataset preprocessed in the same way as outlined in \cref{sec:datasets}), however, we believe that these insights might hold true for a broader set of use-cases.

Our experiments have shown the importance of the hyperparameter $t$, which is responsible for the data scaling and 
whose importance has been emphasized in \cite{canatar2023bandwidth, Shaydulin_2022}. 
This raises an interesting question whether there is a good universal scaling factor for the Hamiltonian evolution feature map. To investigate this, we have found the values of $t$ that provide the best performance in terms of accuracy on average (over all other hyperparameter settings). Additionally, \textit{inner} and \textit{distance} kernels are considered separately as the \textit{distance} kernel has an additional parameter $\gamma$, which exerts control over the scaling as well. As can be seen in \cref{fig:data_scaling}, the datasets are scaled down to a similar range for both types of kernels separately. From here, it follows, that scaling the data down to have standard deviation of approximately $0.49$ for the \textit{inner} kernel and $0.22$ for the \textit{distance} kernel would be a good initial guess.

The findings from the relabeled datasets contrast a universally optimal dataset scaling. As shown in \cref{sec:rela} the optimal scale of the input data would actually be far from the $[-1,1]$ range because the best $t$ value is $t=32$. Therefore, the suggested range will be a good initial guess based on the fact that it was observed for all the original datasets, but there is no guarantee connected with it, especially for datasets that might have an inductive bias that favors a QML model.


Similar to the scaling-discussion above, across many datasets the best values for $t$ are usually within the narrow range $[0.1,1]$ for the \textit{inner} kernel. The \textit{distance} kernel also works well for even smaller $t$ (\cref{fig:app_acc_time_dist_inner}), since this can be compensated for with $\gamma$. The hyperparameter $\gamma$ is correlated with $t$ (\cref{fig:heat_gamma}) and has an optimal range of $[0.1,10]$ (\cref{sec:gamma}). 

In our study, we observed that $T$ had little influence and could be set, for example, to $T=9$ or even smaller to avoid deep circuits. On the other hand, the accuracy was not affected by $K$, but projecting appeared to hinder GD. Based on our findings, $K=6$ would be the optimal selection for our setup. Moreover, this choice is the least computationally expensive, since the number of possible subsystems scales as $N!/(N/2)!^2$ (binomial coefficient), where $N$ is the number of qubits.
\noindent The pipeline described above can be summarized as follows:
\begin{itemize}
    \item Choose $K=N$, $T=9$ and any Haar random initial state;
    \item Explore $t$ in the interval $[0.1,1]$, starting with the one that would scale the data into similar ranges as in \cref{fig:data_scaling};
    \item Choose \textit{inner} kernel to avoid hyperparameter optimization and easier realization; choose \textit{distance} kernel for a better performance at the cost of optimizing $\gamma$ (optimize $\gamma$ and $C$ in a combined fashion);
\end{itemize}

\begin{figure}
  \centering
  \includegraphics[width=\linewidth]{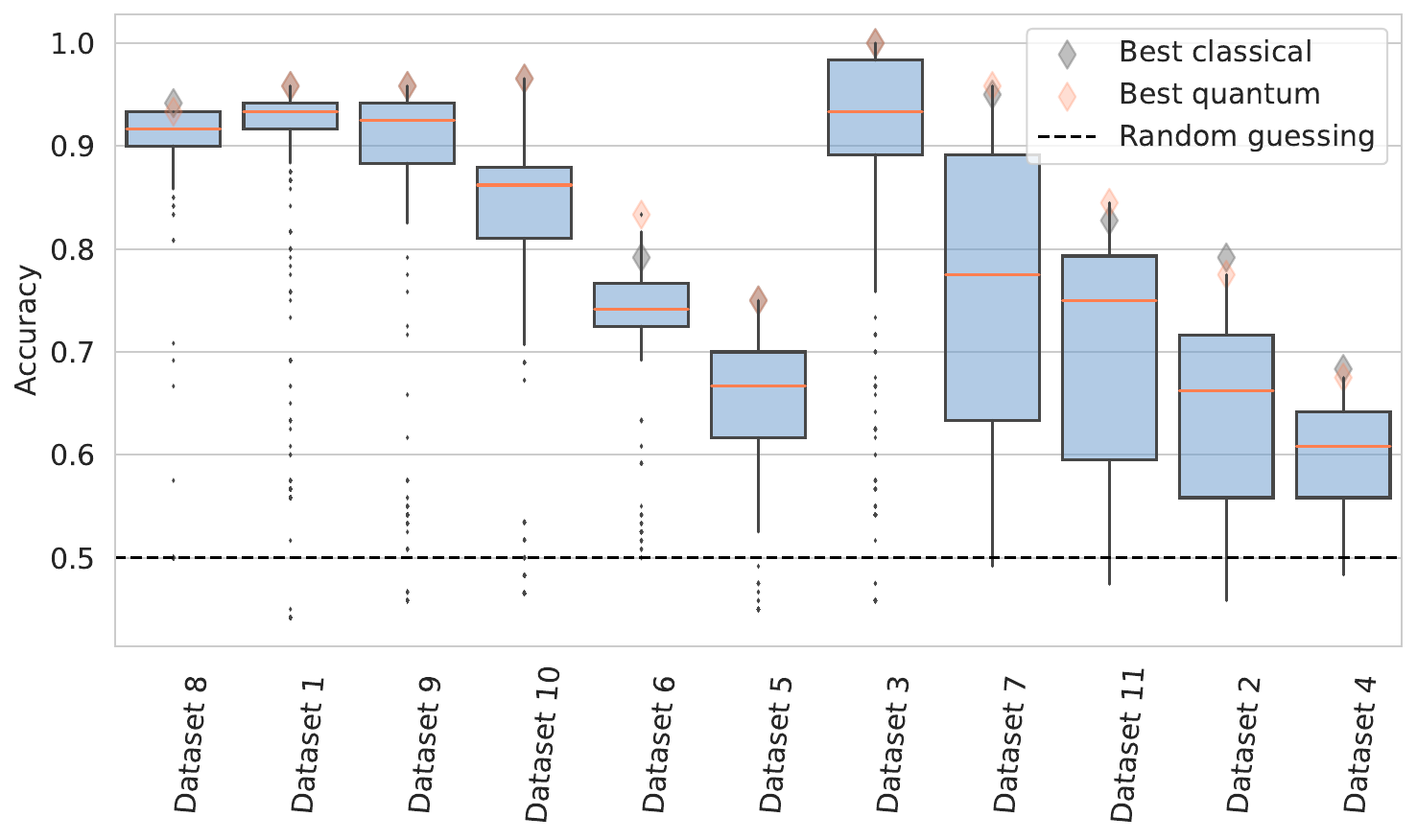}
  \caption{Distribution of accuracies for each dataset (\cref{tab:data}) achieved by an SVM with a quantum kernel for a reduced search grid of hyperparameters. The diamonds indicate the best accuracies achieved by classical and quantum models. The dashed line indicates the accuracy of guessing randomly.}
  \label{fig:acc_over_dataset_pipelined}
\end{figure}

When this pipeline is applied to the datasets in \cref{tab:data} instead of the full grid search of the main experiment (\cref{sec:experiment}), there is only a minor decrease in best quantum performance while the runtime is reduced significantly. The accuracy distribution in the search grid (\cref{fig:acc_over_dataset}) narrows down to the one displayed in \cref{fig:acc_over_dataset_pipelined}. 

\section{Discussion}

The experiment demonstrated that the \textit{distance} kernel outperforms the \textit{inner} kernel when $\gamma$ is optimized, but when we average $\gamma$ over the range we investigated, the opposite is true (\cref{fig:kernel_marginal} and \cref{fig:app_kernel_marginal_gamma_avg}). 
Our results also indicate that the each hyperparameter has compatible importance for both the accuracy and the GD (\cref{fig:importance_barchart}).
These importances, which are extracted from the GBDT model (\cref{sec:analysis}), match our intuition based in the standard deviations in the marginals and the formulas.

The $\gamma$ hyperparameter is easy to optimize in terms of computational resources and because of its single extreme point. Since it can also make or break the \textit{distance} model, it should be optimized. 
$\gamma$ also shows correlations with other parameters, most importantly with $t$. Since both $t$ and $\gamma$ regulate the bandwidth, they are interchangeable to a certain degree. Hence, the behavior of the \textit{distance} kernel over the \textit{inner} kernel when varying $t$ is more robust. This can be seen in \cref{fig:acc_and_gs_heatmap}, which also highlights the fact that the areas with good accuracy and the ones with high GD are rather distinct, especially regarding the dependence on $t$.
The importance of the size of the RDM $K$ is relatively small for the setting we investigated (\cref{fig:importance_barchart}), but we assume that the importance of this parameter to increase with the number of features (qubit count) as the exponential concentration increases in the same way. Notably, the importance of $K$ also increases when we set $\alpha_k = 1$ (\cref{eq:rdm_inner} and \cref{eq:rdm_distance}), which is discussed in \cref{ap:alpha}. This choice however also leads to a stronger dependence of the bandwidth on $K$.
$T$ has a relatively insignificant effect on the performance or the GD as already stated in \cite{Shaydulin_2022}. This suggests that an exact evolution in the Hamiltonian evolution feature map is not necessary, meaning that the approximation one tries to reduce by Trotterization actually does not result in a worse embedding compared to the exact evolution. From the formula stated in Equation (\cref{eq:embedding}), one cannot expect independence from the permutation of the features. In \cref{ap:permutations} we investigate this effect and determine that different permutations lead to different predictions, but that the impact is not significant.

In a supervised learning setup, we can consider on one hand the complexity of the features and on the other hand the complexity of the labels. The complexity of the labels can completely change assumptions based on the features is the reason why the GD and the accuracy sometimes have different dependencies. It is also the reason why a relabeled dataset and the original one have little similarities in terms of accuracy despite having the same features. Understanding the intricate relationship between the characteristics of the features, complexity of the labels, GD metric and empirical difference between quantum and classical learners can help us in the search for real world datasets that favor the inductive bias of quantum KMs. The results from this study provide us with initial hints.
For example, it might in general be better to have a higher $t$ with a lower $\gamma$ compared to a low $t$ with a high $\gamma$. Both possibilities can have the same accuracy, but the first one is closer to the area with a high GD as the second one based on \cref{fig:heat_gamma} and therefore might be a better initial point. Choosing a larger value for $K$ notably increases the GD and slightly enhances performance. This, nonetheless, leads to Hilbert spaces with higher dimensions and convergence of the kernel to a fixed value \cite{thanasilp2022exponential}. RDM kernels might remedy this problem by contracting the space, which comes at an expense of the possible expressivity loss (as indicated by GD). 
Further research is required to understand use-cases that might particularly benefit from the deployment of quantum models.

\section{Conclusion}
To summarize our contributions, in this work we have performed an extensive study of the effects of different hyperparameter settings on the accuracy and the GD score. Our results suggest that even though the hyperparameters have comparable importance for both the accuracy and the GD, the optimal values peaks for the accuracy and GD sometimes differ. This study provides some clues towards understanding the difference between classical and quantum kernel methods.
The most important hyperparameter for both performance metrics is bandwidth $t$, which enforces the claims that were made in previous literature. This hyperparameter controls the scaling of the input data, and among the investigated datasets, the optimal $t$ corresponded to a scaling of the features in a similar range. The GD and the accuracy were observed to depend in an opposing manner on $t$. The \textit{distance} kernel is more robust and performs better than the \textit{inner} kernel thanks to $\gamma$ that can be easily optimized and has a similar effect as $t$. The two alternatives for providing the necessary kernel trace to compute the GD performed almost equally.

Overall, we found optimal ranges for the hyperparameters that work well across the datasets. These findings can be used in order to reduce the computational resources needed when learning a new dataset (\cref{sec:pipeline}). These findings are, however, specific to the size of the feature space (\cref{sec:datasets}) we investigated. One of the main findings of our work concerns $K$, especially since this hyperparameter has no classical equivalence and has not been subject to many experiments yet. We find that it has the most significant difference in importance for the accuracy and the GD. 
How this difference changes with higher dimensions of the feature space is left for future investigations.

Relabeled datasets that posses labels that are more favourable to quantum learners show different dependencies and optimal values than the real world datasets. Some findings, like an optimal range for feature scaling, can break down for them. Depending on the choice of prefactors $\alpha_k$ in \cref{eq:rdm_inner} and \cref{eq:rdm_distance}, the observations for relabeling with different $K$ may change. Specifically, setting $\alpha_k = 1/N_K$ yields greater separation with higher GD, whereas $\alpha_k = 1$ did not produce similar results in our conducted experiments. 
More studies are necessary to make concussive statements about this connection.

In order to further increase the understanding of the GD we also suggest that more experiments with relabeled datasets, different embeddings and choices of RDM kernels are conducted. Further analyzing the spectral properties of the quantum kernel can shed light on the inductive bias of the model \cite{kubler2021inductive, Slattery_2023}. Another interesting future work in context of the quantum Hamiltonian evolution feature map would be to include noise in such a broad hyperparameter analysis, which is ideally done with real hardware runs. Both noisy runs and increasing the number of qubits may provide new insights, particularly regarding the role of $K$,  as exponential concentration \cite{thanasilp2022exponential} would become more pronounced. This hyperparameter governs the size of the projection that helps counteract the effects of the exponential concentration, and, hence, its importance could rise once we consider scenarios that are more affected by this phenomenon.

Given the relationship of QKMs to other QML approaches \cite{schuld2021supervised}, an interesting further research direction would be to investigate how our findings translate to other models. For example, hyperparameters responsible for scaling of the input, such as $t$ and $\gamma$, have been identified to have a dominating influence while their optimal values put the data ranges (see \cref{fig:data_scaling}) way below the industrial standard. This begs the question whether other QML models, such as Quantum Neural Networks, might benefit from a change in the standard data preprocessing procedure to fit the preprocessed data into the identified ranges. 

\section*{Acknowledgements}
The research is part of the Munich Quantum Valley, which is supported by the Bavarian state government with funds from the Hightech Agenda Bayern Plus.

\bibliographystyle{plain} 
\bibliography{main} 

\begin{appendices}
\section{Inner normalized kernel function}\label{app:inner_norm}

Additionally to the kernels considered in the main text, we tested an adaptation of the \textit{inner product based} kernel. The motivation is that the kernel entries can become relatively small for large dimensions and highly entangled states. On the main diagonal the values are related to the purity of a quantum state $Tr(\rho^2)$, which can become $\frac{1}{d}$ where $d$ is the dimension of the Hilbert space. The \textit{normalized inner product based} kernel is defined as

\begin{align}\label{eq:rdm_inner_norm}
        k(x, x') &= \sum^{N_K}_k Tr(\Tilde{\rho}^{(q_k)}(x)\Tilde{\rho}^{(q_k)}(x'))\\  \Tilde{\rho}^{(q_k)}(x) & \coloneqq \frac{\rho^{(q_k)}(x)}{\sqrt{\sum^{N_K}_kTr(\rho^{(q_k)}(x)^2)}}
\end{align}

and therefore remains a valid kernel, because the same feature vectors as in \textit{Nakaji et al.}~\cite{nakaji2022deterministic} can be found for the redefined density matrices (DMs). This normalization affects all entries in the gram matrix, in particular the main diagonal now consists of just ones, as it is common for kernels in general. Important to note is that this kernel is only meaningful in a simulation environment.

\begin{figure}
  \centering
  \includegraphics[width=\linewidth]{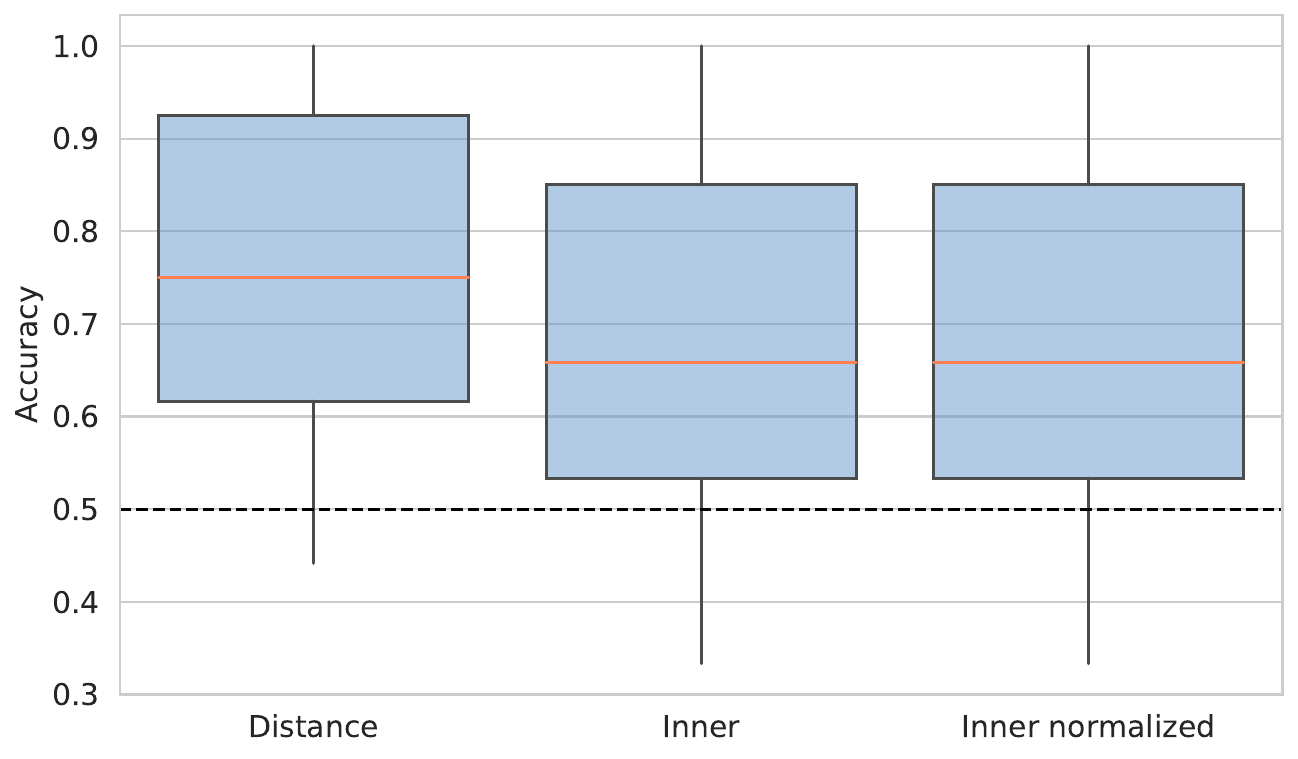}
  \caption{Dependence of the test accuracy on the choice of \textit{basis} for the kernel function. For this comparison $\gamma$ is optimized and not averaged. The specific form of a \textit{distance}-, \textit{inner}- or \textit{inner normalized} based kernel function is given in \cref{sec:quantum_kernels}. The dashed line indicates the accuracy that would be achieved by random guessing.}
  \label{fig:kernel_marginal_with_norm}
\end{figure}

When comparing methods \textit{inner} and \textit{inner normalized} we notice almost no difference in performance (as seen in \cref{fig:kernel_marginal_with_norm}) or GD (as seen in later sections). This implies that this adaptation has almost no influence on classification of the SVM.

\section{Analysing different kernel functions}\label{ap:performances_by_models}

\begin{figure}[H]
    \centering
    \includegraphics[width=\linewidth]{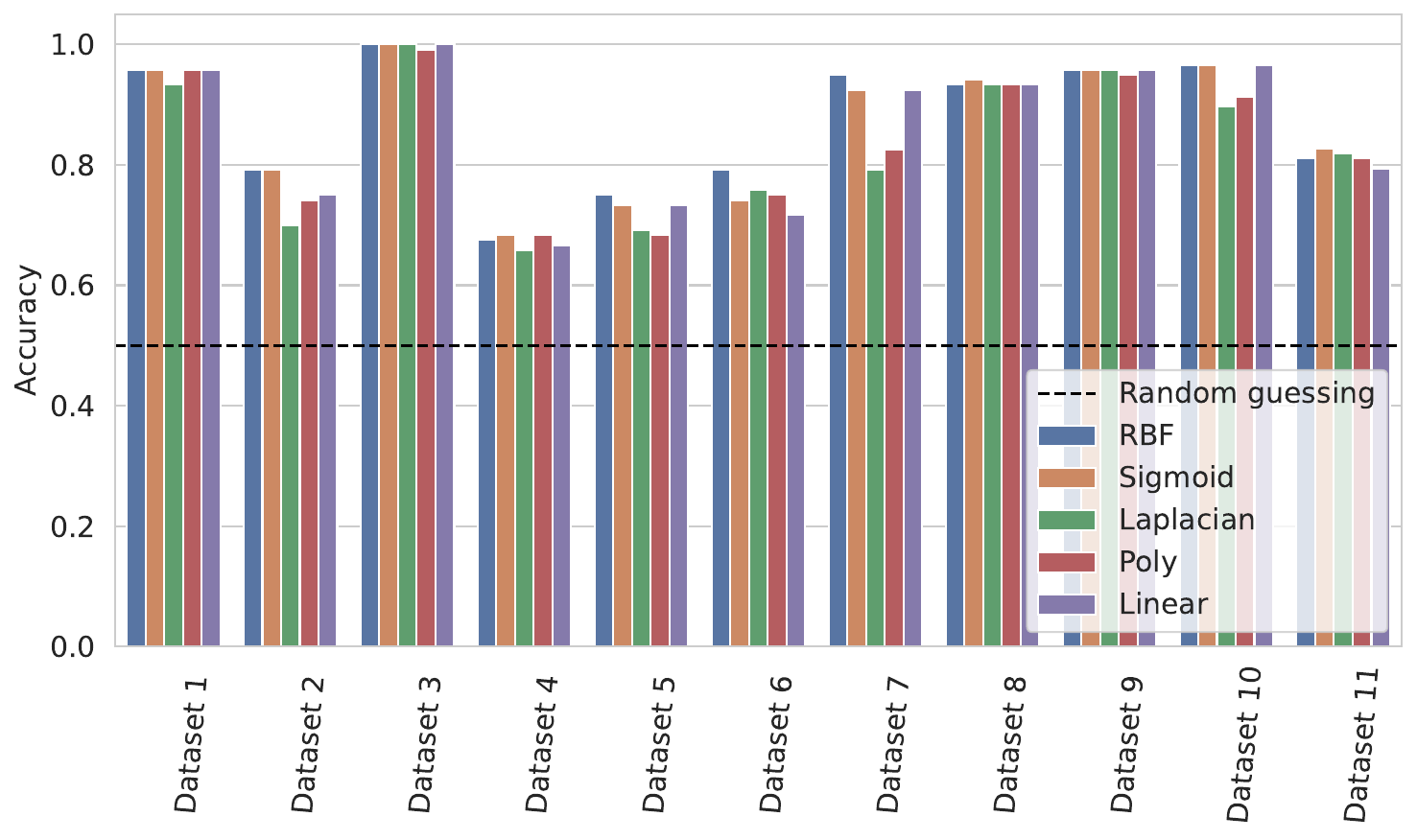}
    \caption{Best achieved test accuracies on each dataset by classical kernels. The dashed line indicates the accuracy that would be achieved by random guessing.}
    \label{fig:best_classical}
\end{figure}

\begin{figure}
    \centering
    \includegraphics[width=\linewidth]{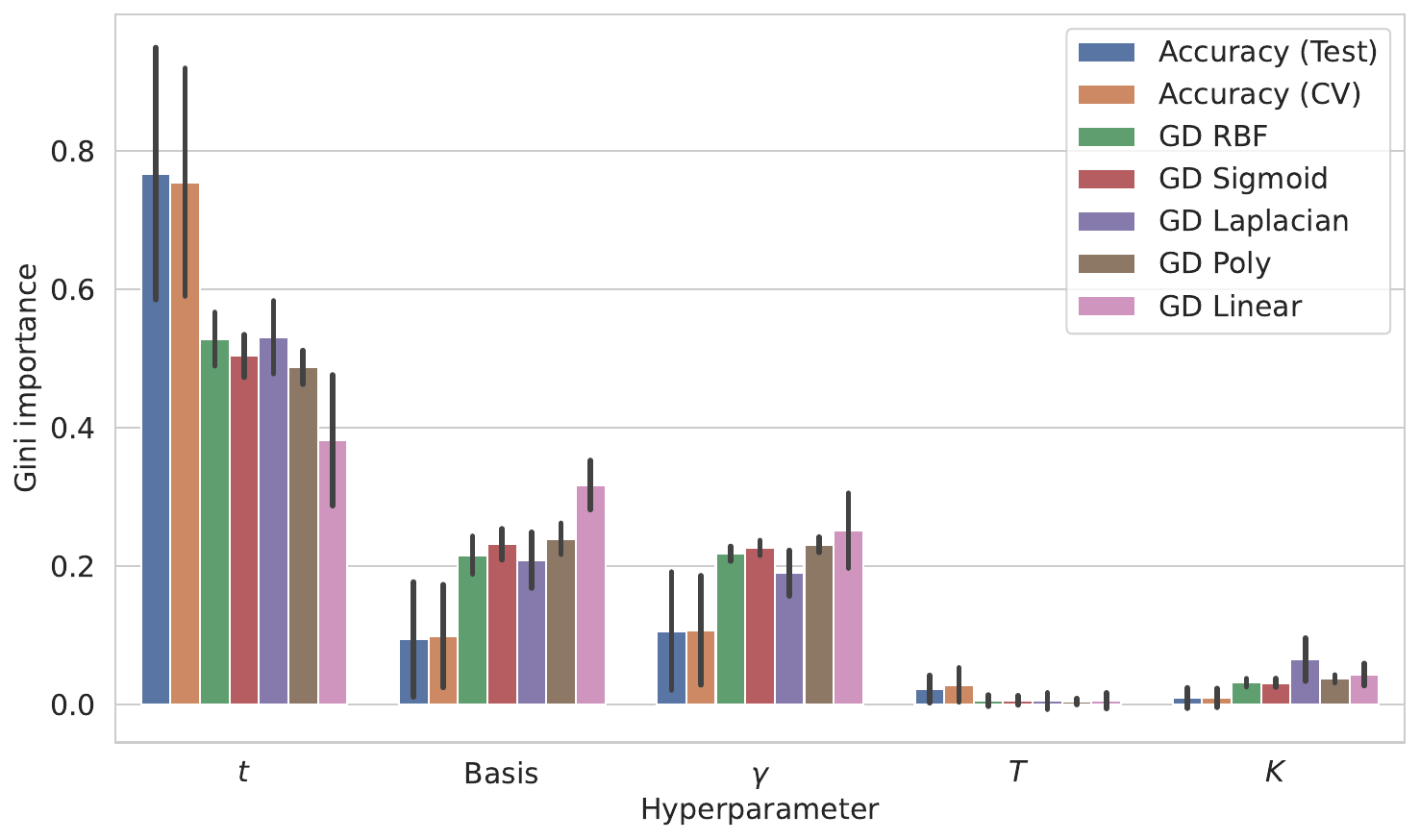}
    \caption{Importance of each hyperparameter for a collection of metrics. These include the test accuracies and cross validation accuracies as well as the GD to 5 different classical kernels. The errorbars illustrate the standard deviations when averaging the importances across the 11 datasets we considered in our study.}
    \label{fig:Gini_all_metrics}
\end{figure}

To validate our choice of a classical RBF kernel as a benchmark and to investigate how this choice affected GD results, we test other standard kernel functions, such as Linear, Polynomial, Sigmoid and Laplacian, in the same setup. \cref{fig:best_classical} portraits the best achieved performances by all models on all available datasets. The RBF performs best on average by a small margin. 
The importance of hyperparameter changes by a small margin as we changes a classical basis of a GD. The importance values shown in \cref{fig:Gini_all_metrics} indicate that the dependence of the GD on the different parameters is relatively universal over the choice of the classical kernel to which this GD is taken.

\begin{figure}[h]
    \centering
    \includegraphics[width=\linewidth]{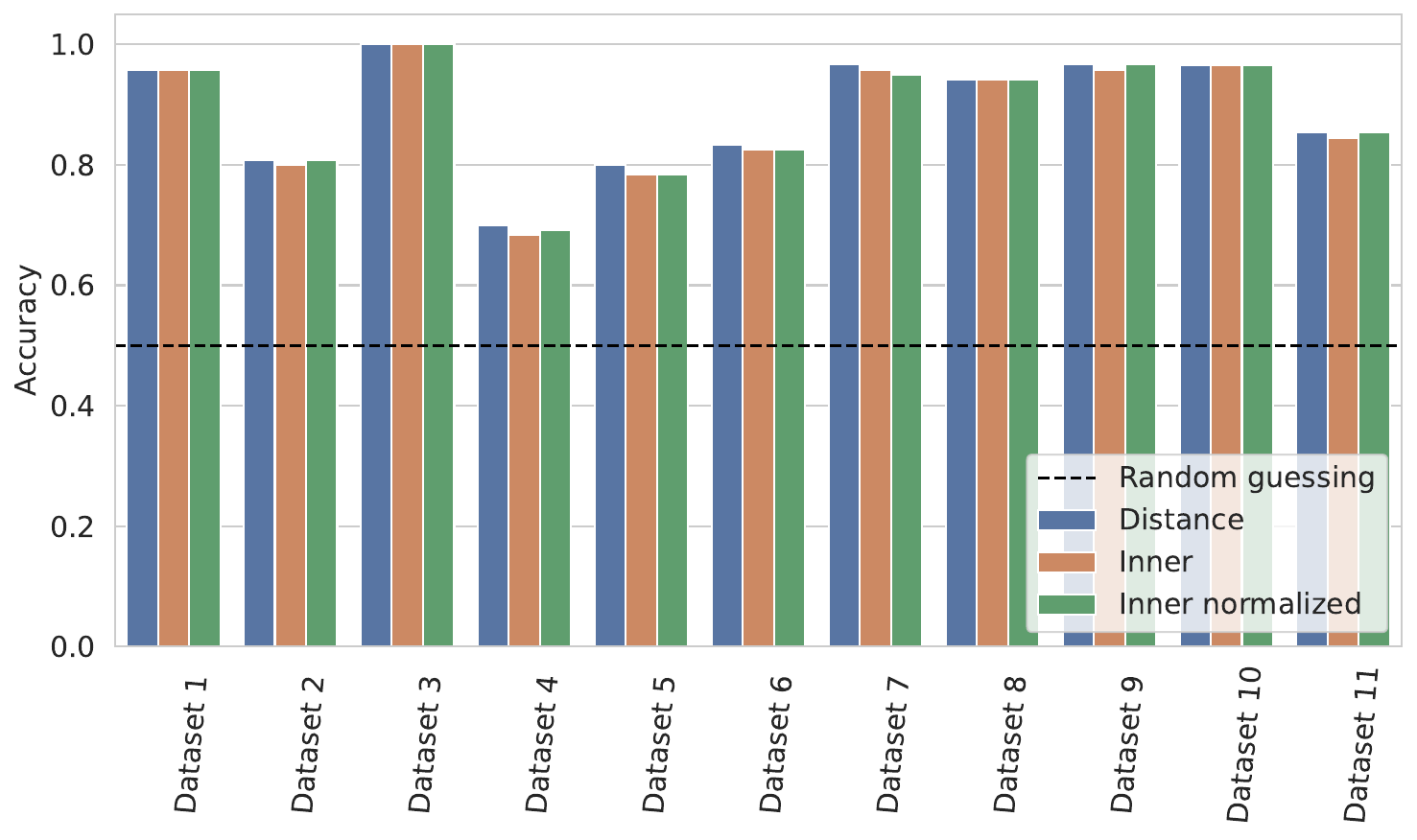}
    \caption{Best achieved test accuracies on each dataset by quantum kernels. The maxima of all three models align very well, with the \textit{distance} kernels being the best performing quantum kernel by a small margin. When looking at the average (\cref{fig:kernel_marginal}) instead of the maximum accuracy, this becomes even more clear. The \textit{inner} and the \textit{inner normalized} align closely. The dashed line indicates the accuracy of guessing randomly.}
    \label{fig:best_quantum}
\end{figure}

To illustrate that the implications of \cref{fig:kernel_marginal} persist for every dataset individually with little exceptions, we show in \cref{fig:best_quantum} the individual best performance of these three quantum methods (\textit{distance}, \textit{inner} and \textit{normalized inner}) for each dataset. This plot illustrates a slight advantage of the \textit{distance} kernel that was already determined in the original plot.
Additionally, in both \cref{fig:kernel_marginal} and \cref{fig:best_quantum} we see that the maxima of the accuracies of the \textit{inner} and the \textit{inner normalized} align well.

\section{Additional details of empirical study}\label{ap:addition}
\begin{figure}
  \centering
  \includegraphics[width=\linewidth]{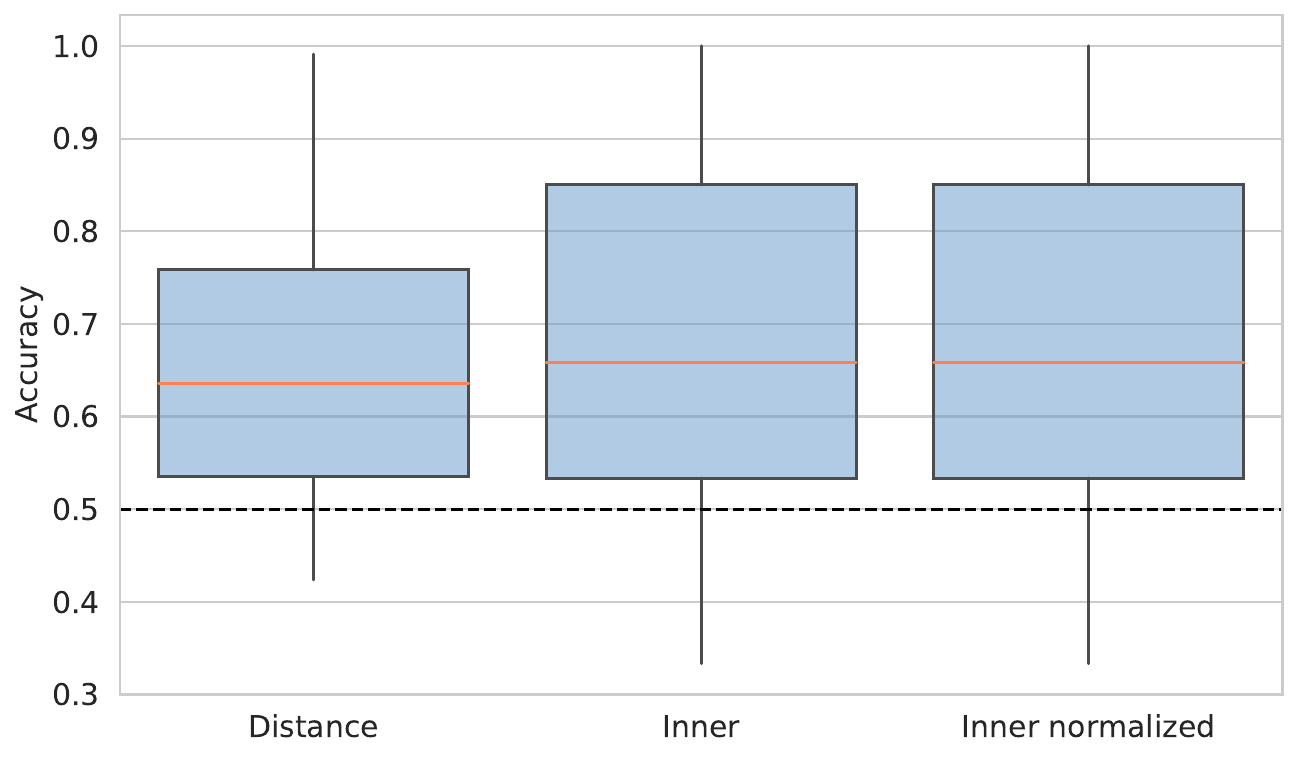}
  \caption{Dependence of the test accuracy on the choice of \textit{basis} for the kernel function. In contrast to \cref{fig:kernel_marginal} $\gamma$ is averaged over too along with all the other parameters except for \textit{basis}. The specific form of a \textit{distance}-, \textit{inner}- or \textit{inner normalized} based kernel function is given in \cref{sec:quantum_kernels}. The dashed line indicates the accuracy of guessing randomly.}
  \label{fig:app_kernel_marginal_gamma_avg}
\end{figure}

\begin{figure}
  \centering
  \includegraphics[width=\linewidth]{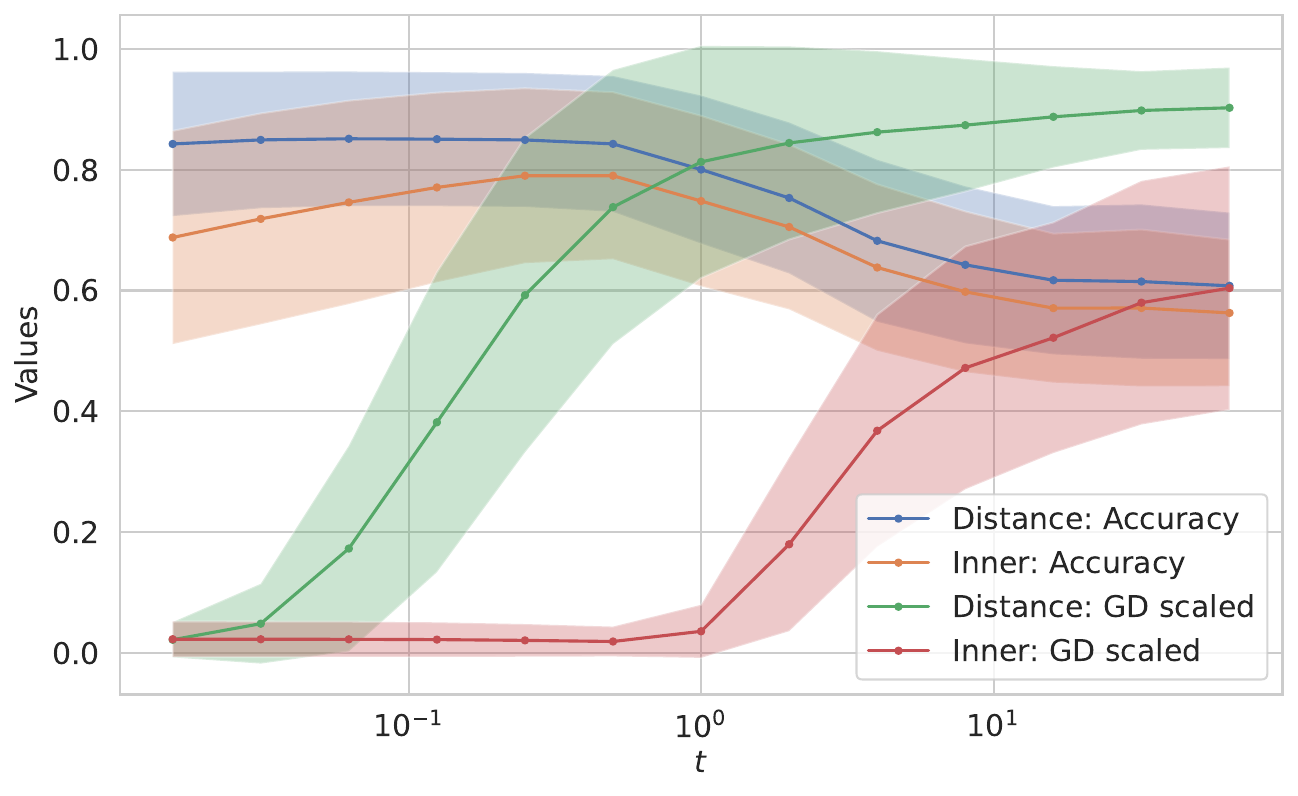}
  \caption{Dependence of the test accuracy and the GD on $t$ (evolution time) for the \textit{distance} and the \textit{inner} kernel separately.}
  \label{fig:app_acc_time_dist_inner}
\end{figure}

In this section, additional figures are shown to support statements from \cref{sec:results}. The marginal plots for each dataset show the mean over all other parameter settings as already described in \cref{sec:analysis}. \cref{fig:app_acc_time_dist_inner} has the means to highlight the difference between the \textit{distance} and the \textit{inner} kernel with focus on the stability of the \textit{distance} one for small $t$ values. \cref{fig:app_kernel_marginal_gamma_avg} also compares the different choices of \textit{basis}, but for an average over $\gamma$ in contrast to \cref{fig:kernel_marginal} where this parameter is optimized. The remaining figures in this section have the purpose to display common or different behaviour of the parameter dependencies across the different datasets.
\begin{figure}
  \centering
  \includegraphics[width=\linewidth]{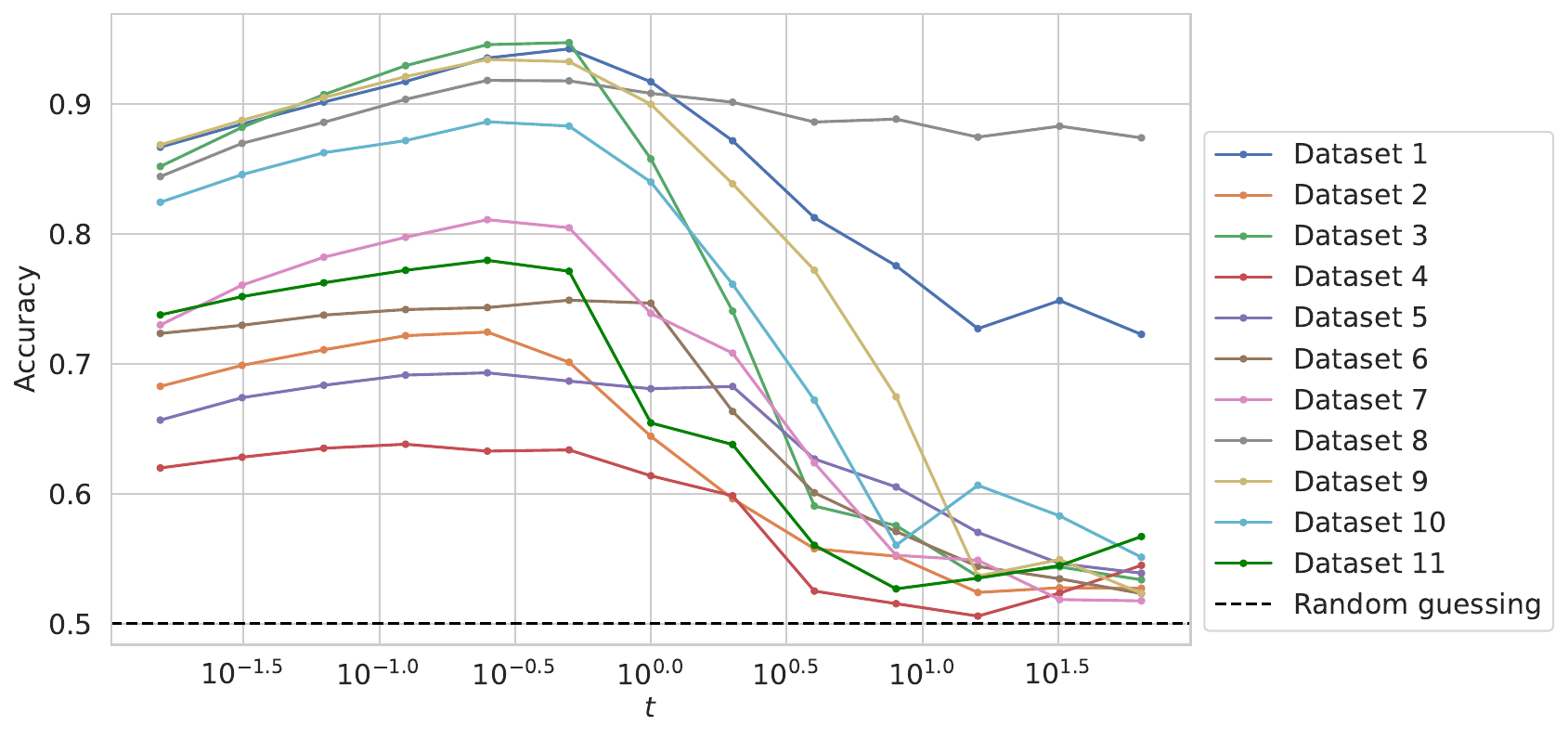}
  \caption{Dependence of the test accuracy on $t$ (evolution time) for each of the 11 original datasets listed in \cref{tab:data} individually.}
  \label{fig:app_acc_time}
\end{figure}
\begin{figure}
  \centering
  \includegraphics[width=\linewidth]{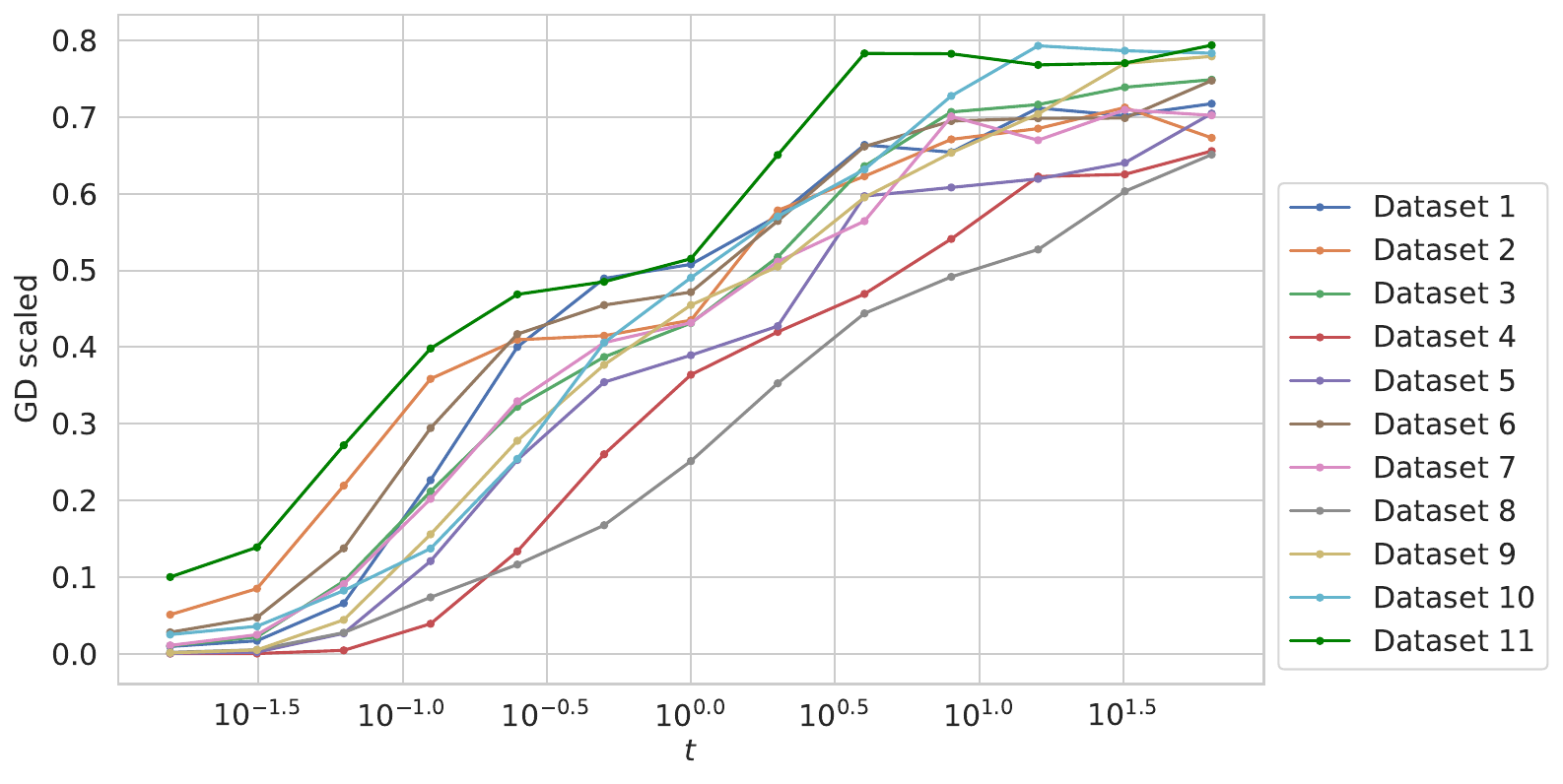}
  \caption{Dependence of the GD on $t$ (evolution time) for each of the 11 original datasets listed in \cref{tab:data} individually.}
  \label{fig:app_GD_time}
\end{figure}
\begin{figure}
  \centering
  \includegraphics[width=\linewidth]{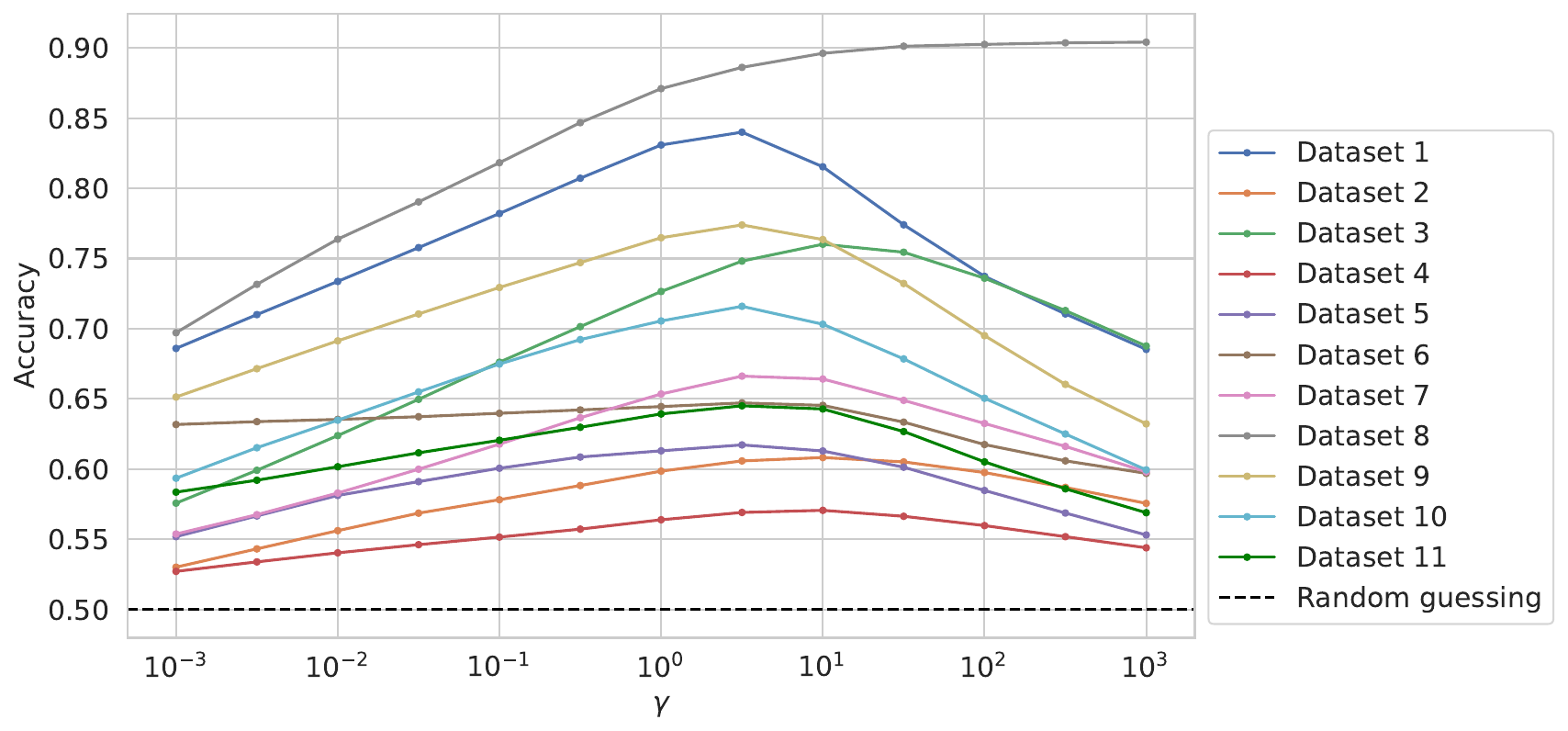}
  \caption{Dependence of the test accuracy on $\gamma$ (bandwidth) for each of the 11 original datasets listed in \cref{tab:data} individually.}
  \label{fig:app_acc_gamma}
\end{figure}
\begin{figure}
  \centering
  \includegraphics[width=\linewidth]{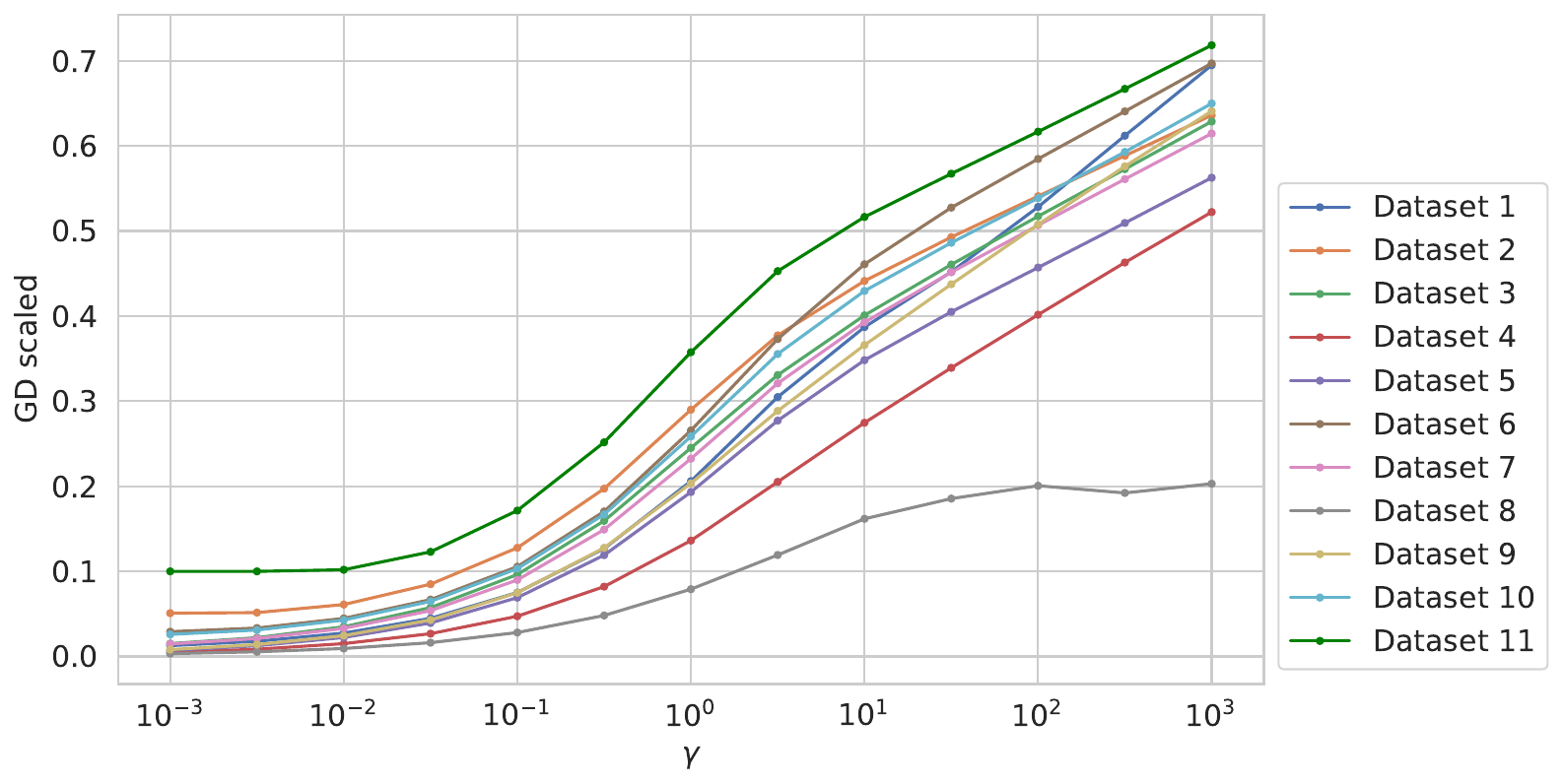}
  \caption{Dependence of the GD on $\gamma$ (bandwidth) for each of the 11 original datasets listed in \cref{tab:data} individually.}
  \label{fig:app_GD_gamma}
\end{figure}
\begin{figure}
  \centering
  \includegraphics[width=\linewidth]{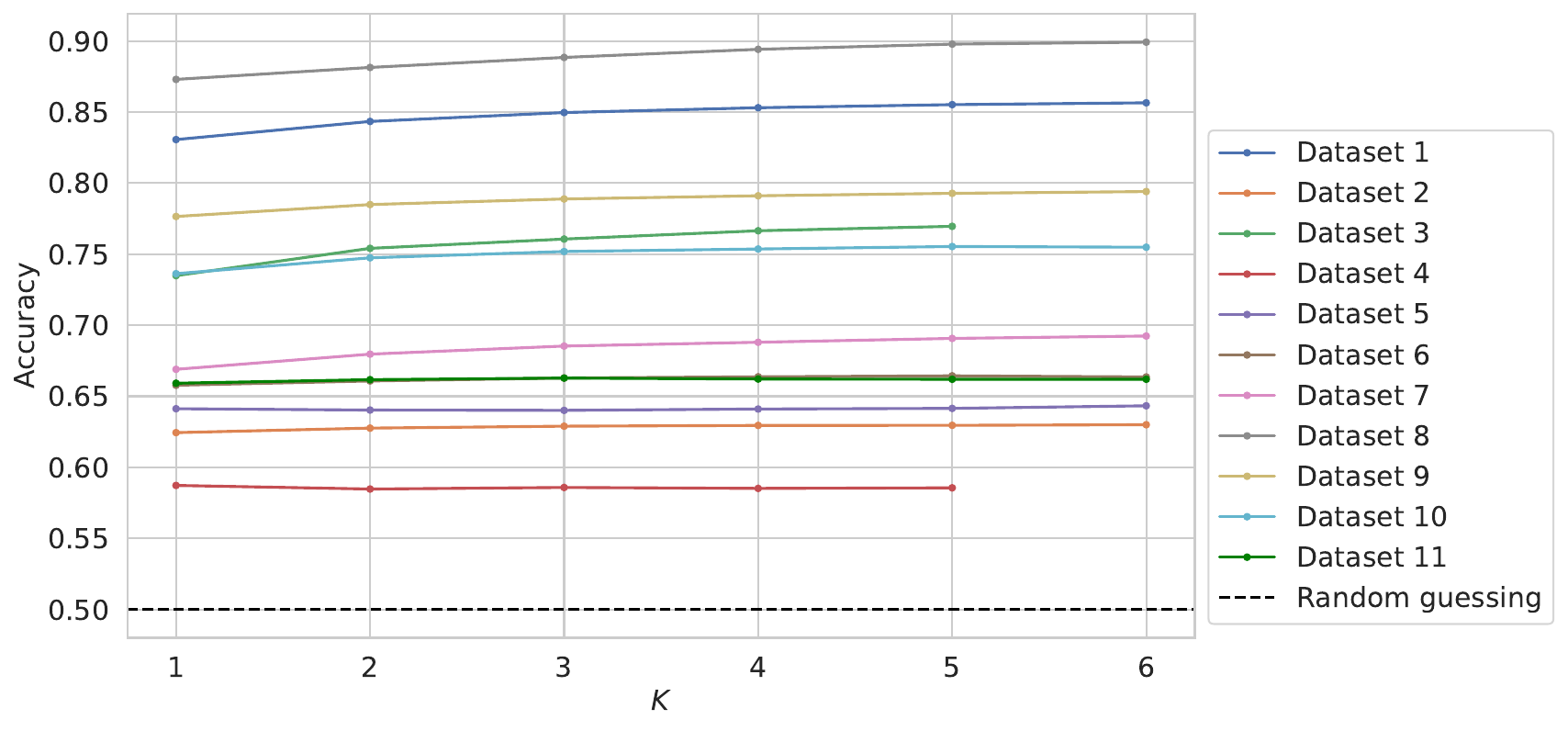}
  \caption{Dependence of the test accuracy on $K$ (RDM size) for each of the 11 original datasets listed in \cref{tab:data} individually.}
  \label{fig:app_acc_K}
\end{figure}
\begin{figure}
  \centering
  \includegraphics[width=\linewidth]{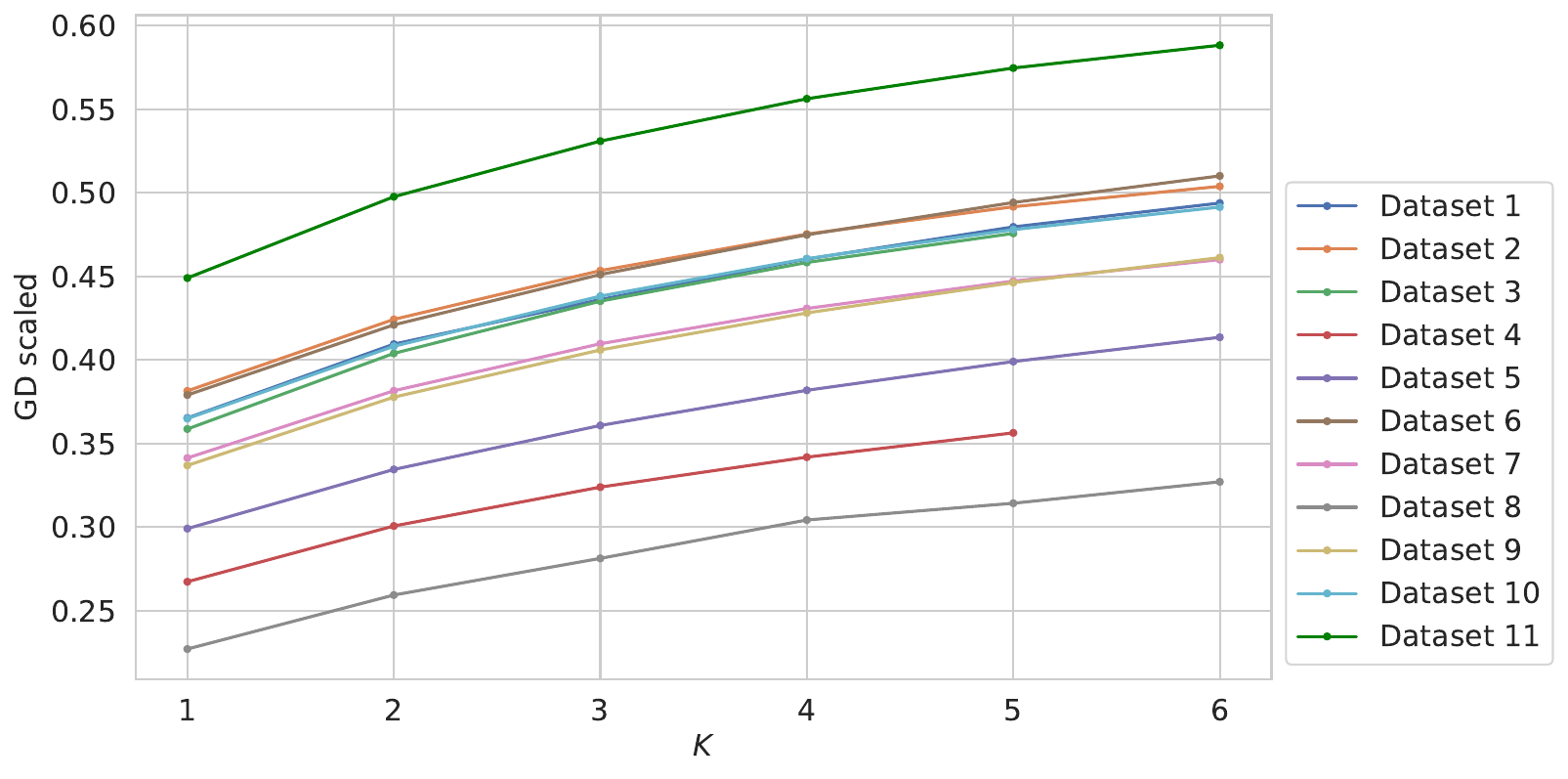}
  \caption{Dependence of the GD on $K$ (RDM size) for each of the 11 original datasets listed in \cref{tab:data} individually.}
  \label{fig:app_GD_K}
\end{figure}
\begin{figure}
  \centering
  \includegraphics[width=\linewidth]{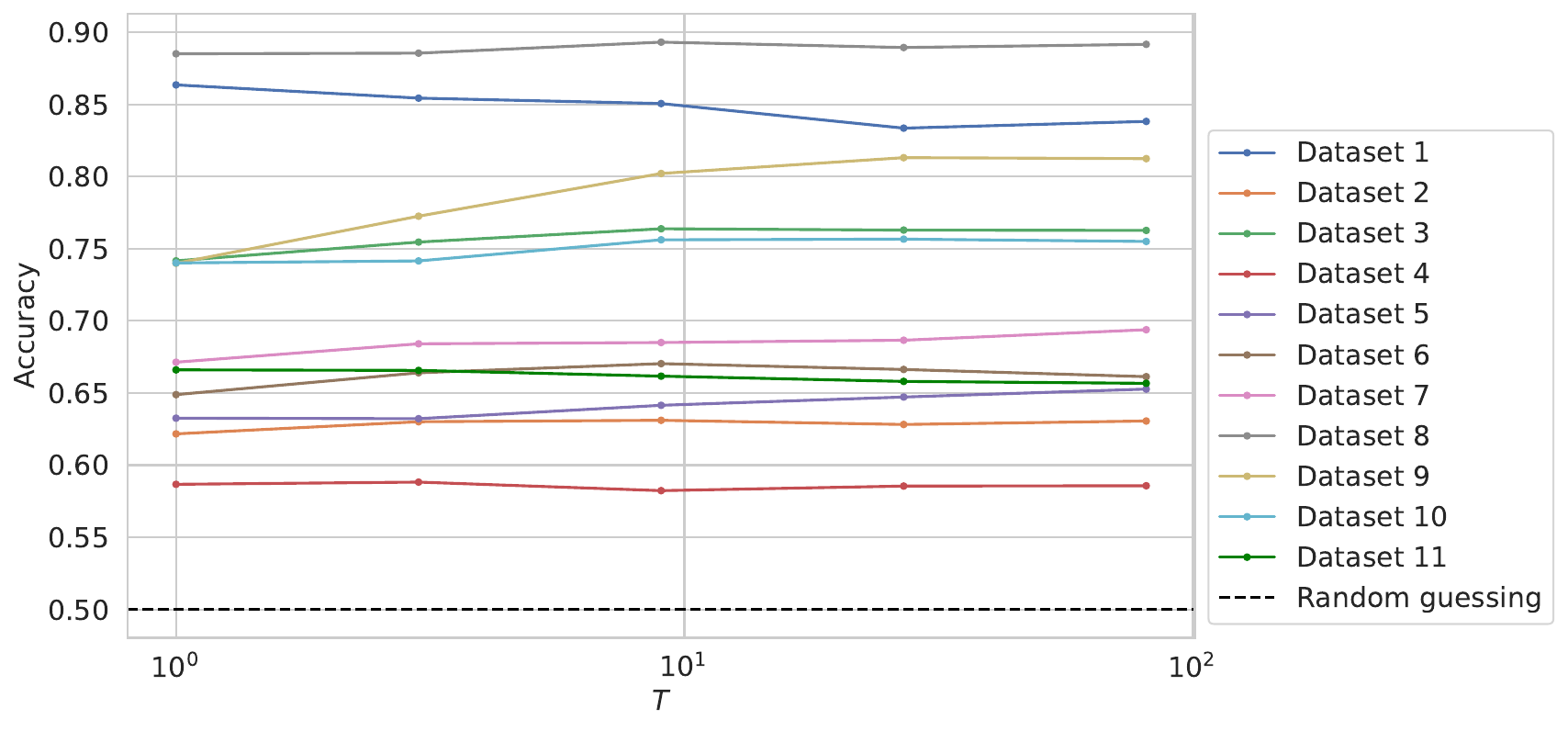}
  \caption{Dependence of the test accuracy on $T$ (number of Trotterization steps) for each of the 11 original datasets listed in \cref{tab:data} individually.}
  \label{fig:app_acc_T}
\end{figure}
\begin{figure}
  \centering
  \includegraphics[width=\linewidth]{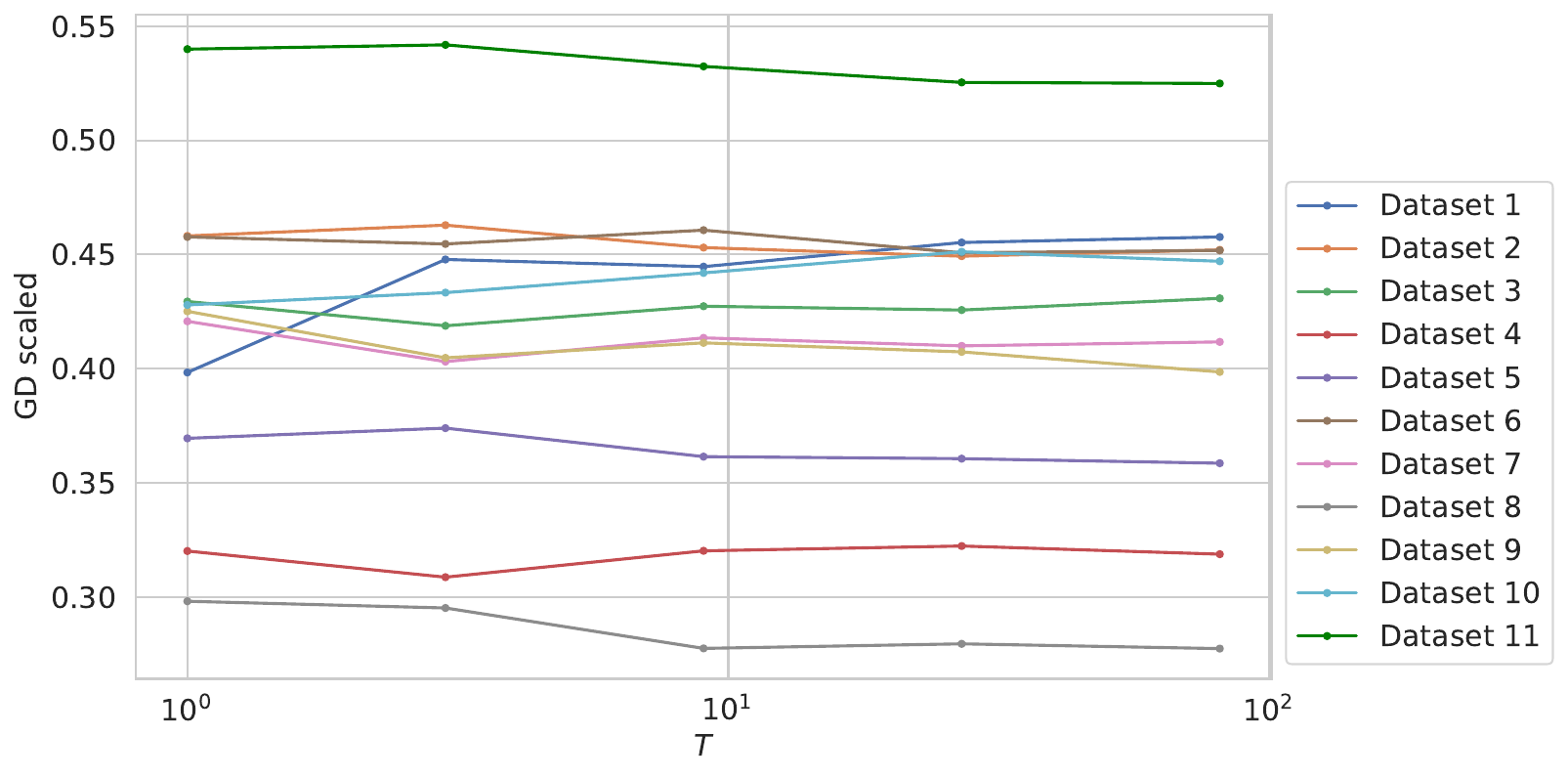}
  \caption{Dependence of the GD on $T$ (number of Trotterization steps) for each of the 11 original datasets listed in \cref{tab:data} individually.}
  \label{fig:app_GD_T}
\end{figure}
\begin{figure}
  \centering
  \includegraphics[width=\linewidth]{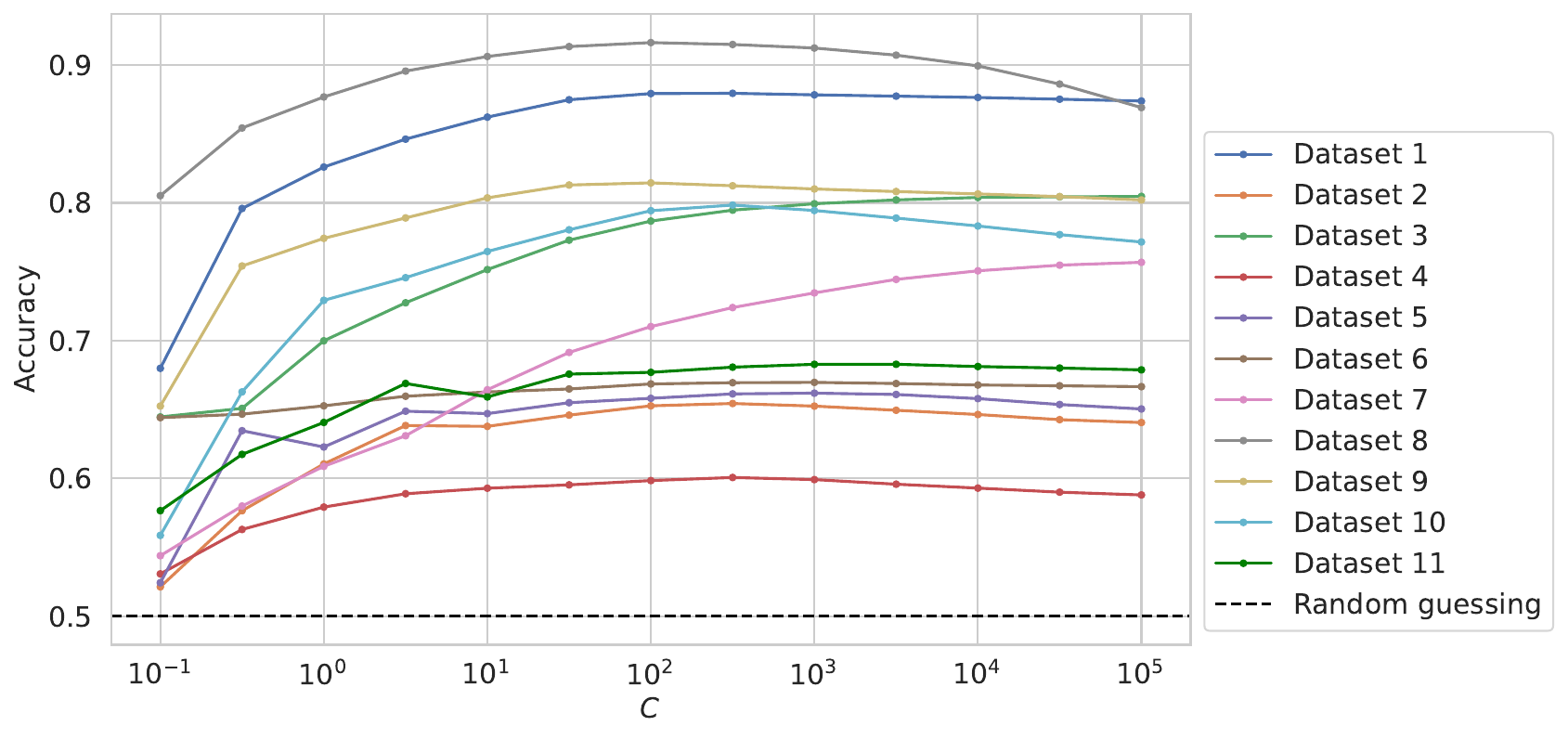}
  \caption{Dependence of the test accuracy on $C$ (regularization strength) for each of the 11 original datasets listed in \cref{tab:data}.}
  \label{fig:app_acc_C}
\end{figure}

\section{Effects of changing dataset preprocessing steps}\label{ap:preprocessing}
\begin{figure}
  \centering
  \includegraphics[width=\linewidth]{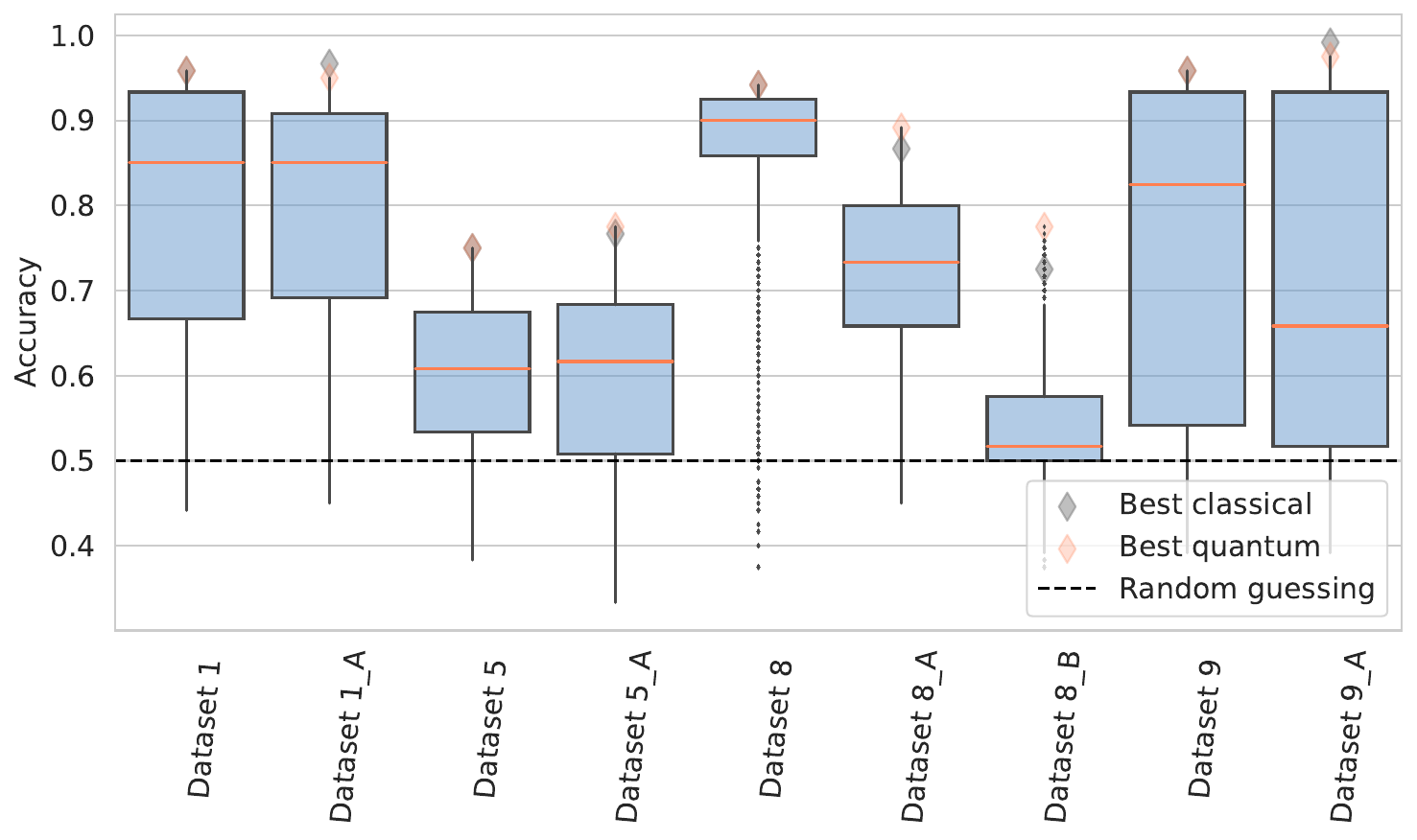}
  \caption{Distribution of accuracies achieved by an SVM with a quantum kernel on the test set for all hyperparameters setting given in \cref{tab:hyperparameters}. The different boxplots correspond to different prepossessings. Additionally, the diamonds indicate the best accuracies achieved by classical and quantum models. The dashed line indicates the accuracy of guessing randomly.}
  \label{fig:acc_over_dataset_prepro}
\end{figure}

\begin{figure}
  \centering
  \includegraphics[width=\linewidth]{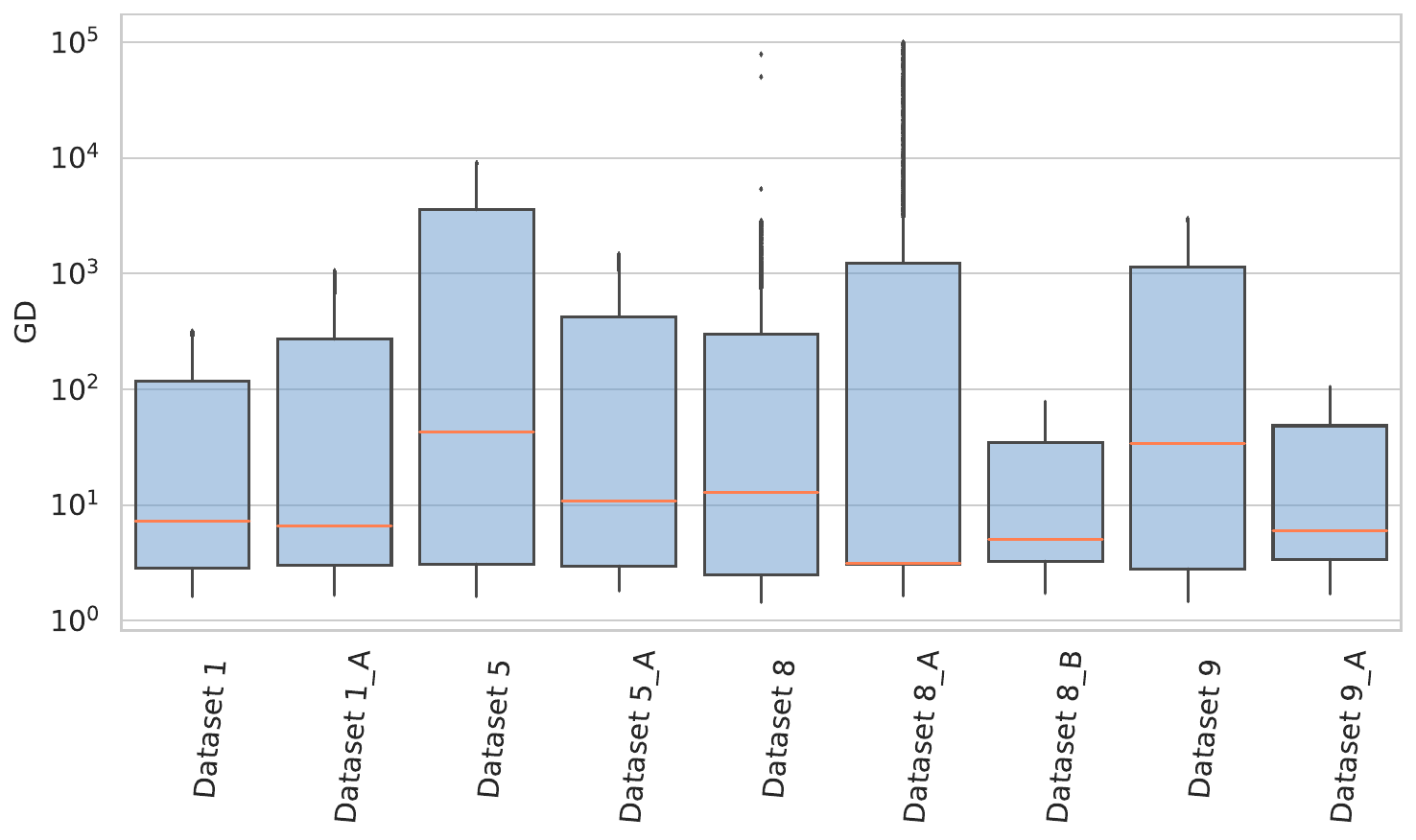}
  \caption{Distribution of GD between a quantum kernel and the classical RBF kernel over all hyperparameters setting given in \cref{tab:hyperparameters}. The different boxplots correspond to different prepossessings of datasets 1, 5, 8 and 9 and are cut off at $10^5$ ($\approx1.2\%$ cut off).}
  \label{fig:GD_over_dataset_prepro}
\end{figure}

In addition to the preprocessing routine described in \cref{sec:datasets} we investigated how keeping or dropping correlated features affects the accuracies. For the whole discussion in the main part of this work highly correlated features were kept. In this section \textit{\_A} indicates that any feature, where the correlation with another feature was greater than 0.75 was eliminated. This step affects only datasets 1, 5, 8 and 9. For dataset 8 we also tested the preprocessing steps from \cite{TF_quantum} (indicated as \textit{\_B}) meaning that the datapoints are scaled to the range $[0,1]$ and principal component analysis  (PCA) was used for dimensionality reduction. We restrict it to just one random initial state for this section. \cref{fig:acc_over_dataset_prepro} and \cref{fig:GD_over_dataset_prepro} demonstrate that different preprocessing techniques have the greatest impact on dataset 8, where they lead to better results. This particular dataset is unique, as it is the only one based on image data (and therefore has 784 initial features), compared to an initial number of features of 9, 10, and 30 for datasets 1, 5, and 9, respectively. Furthermore, the sample size for this dataset is substantially larger than the other datasets.

\section{Effects of feature permutations}\label{ap:permutations}
\begin{figure}[H]
  \centering
  \includegraphics[width=\linewidth]{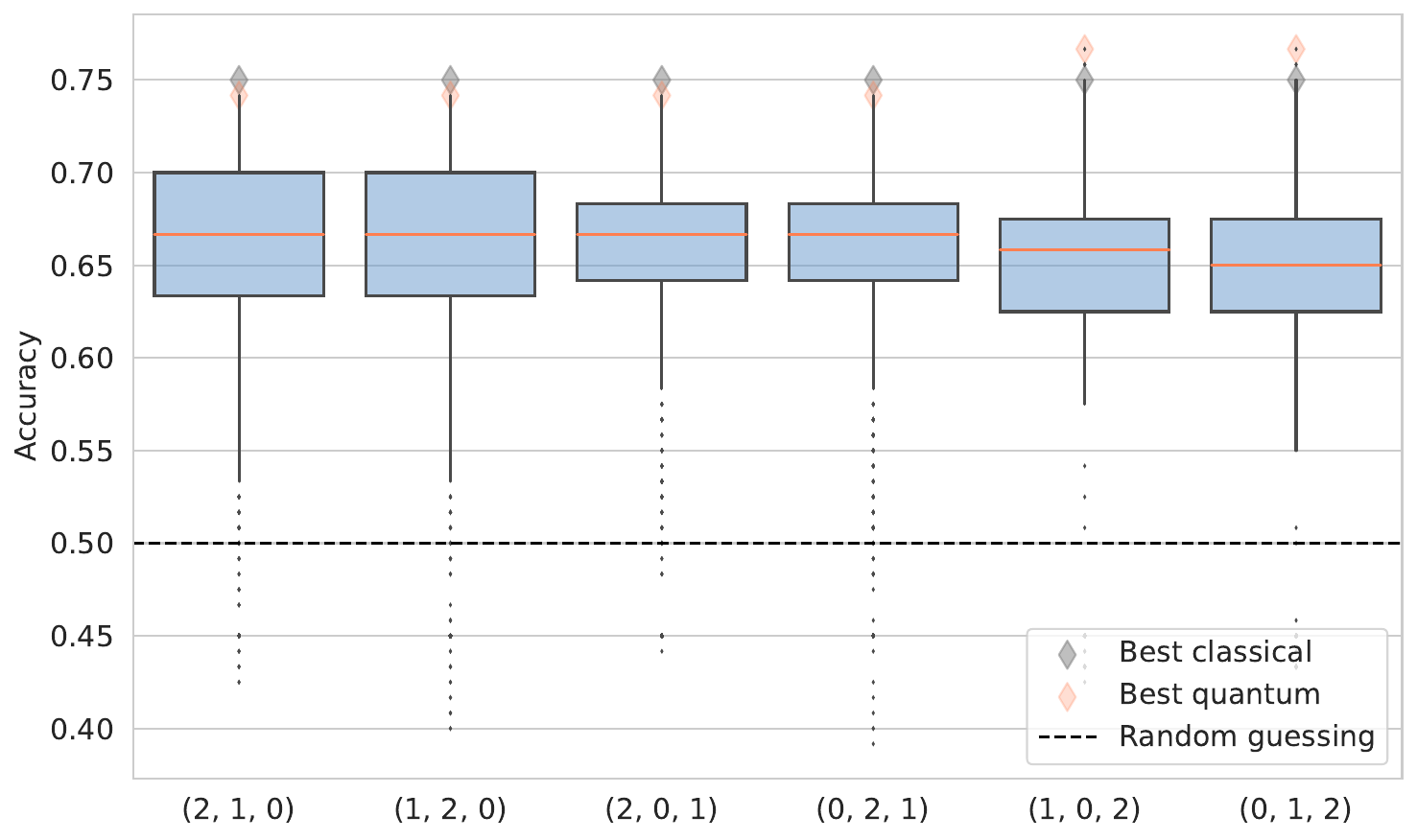}
  \caption{Distribution of accuracies for dataset 5 with 3 features across different permutations. The dashed line indicates the accuracy of guessing randomly.}
  \label{fig:perm}
\end{figure}

In the quantum Hamiltonian evolution feature map given in \cref{eq:embedding}, each feature acts on the qubit with the same index as the feature and the qubit with this index+1. Thus, unlike most classical kernels, different permutations of the input features do not lead to exactly the same model. Moreover, the last feature is entangled with an additional qubit and therefore depends on only one feature directly. The first qubit also depends on only one feature, while the other qubits depend on two. The optimal permutation is difficult to optimize, so we attempt to gain some insight into the severity of this effect.
To this end, we apply the pipeline of \cref{sec:pipeline} to a selection of datasets for each possible permutation of the features. Because the number of possible permutations $N_{p}$ of the number of features $N_{f}$ grows like $N_{p} = N_{f}!$, we restrict to just 3 features for this experiment. For each possible hyperparameter setting, we compute the standard deviation of test accuracy across permutations. These standard deviations are then averaged across settings and listed in \cref{tab:perm} along with the mean test accuracy across all settings and permutations. The values are reported for the GD and the test accuracy. For the specific choice of 3 features and for the datasets we consider, we did not find a noticeable effect on GD, but not much effect on test accuracy (see also \cref{fig:perm}).

\begin{table}
\centering
\begin{tabular}{lcccc} 
\toprule
\textbf{ID} & \textbf{mean accuracy} & \textbf{std accuracy} & \textbf{mean GD} & \textbf{std GD}\\
\midrule
\multirow{1}{*}{Dataset 1} & 0.87 & 0.01 & 61.82 & 5.73\\
\multirow{1}{*}{Dataset 2} & 0.66 & 0.03 & 220.61 & 14.25\\
\multirow{1}{*}{Dataset 3} & 0.85 & 0.03 & 103.17 & 12.91\\
\multirow{1}{*}{Dataset 4} & 0.59 & 0.02 & 275653.97 & 75473.20\\
\multirow{1}{*}{Dataset 5} & 0.65 & 0.02 & 94790.87 & 32293.60\\
\multirow{1}{*}{Dataset 6} & 0.72 & 0.02 & 382.86 & 66.67\\
\multirow{1}{*}{Dataset 7} & 0.61 & 0.03 & 986.72 & 227.56\\
\multirow{1}{*}{Dataset 8} & 0.88 & 0.01 & 685.82 & 30.94\\
\multirow{1}{*}{Dataset 9} & 0.87 & 0.02 & 26158.52 & 7031.87\\
\multirow{1}{*}{Dataset 10} & 0.78 & 0.06 & 389.69 & 130.32\\
\multirow{1}{*}{Dataset 11} & 0.72 & 0.04 & 200.71 & 43.88\\
\bottomrule
\end{tabular}
\caption{List of datasets and the effects of permuting the features. \textit{mean} denotes the mean value across all hyperparameter settings and permutations and provides context for\textit{std}. \textit{std} denotes the standard deviation across the permutations, from which we then take the mean over all hyperparameter settings. This is given for the test accuracy and the GD.}
\label{tab:perm}
\end{table}

\section{Effects of $\alpha_K$ parameter}\label{ap:alpha}
Both $\alpha_K=1/N_K$ and $\alpha_K=1$ in equations \cref{eq:rdm_inner} and \cref{eq:rdm_distance} result in valid kernels, but we find that the latter choice leads to counter-intuitive observations especially regarding the GD. This choice mainly affects observations based on different values of the hyperparameter $K$.
For example, \ref{tab:rela_alpha_1} is equivalent to \ref{tab:rela}, but for $\alpha_K=1$ instead of $\alpha_K=1/N_K$. When we make this choice, we see that relabeling for a larger GD does not necessarily lead to a larger separation between the best classical and the best quantum model in terms of test accuracy. Going from $t=0.5$ to $t=32$ increases the GD and the separation for both $\alpha_k$. For different $K$ this is not the case.

For $\alpha_K=1$, the sum over all possible subsystems yields kernel entries that are roughly $\propto N_K$. This suggests an added scaling factor of the bandwidth, leading to strong correlations between $\gamma$ and $K$, as well as $t$ and $K$, which do not exist for $\alpha_K=1/N_K$. Based on this observation, we conclude that $\alpha_K=1/N_K$ is a better choice for this parameter and use it in the main part of this study.

\begin{table}
\centering
\begin{tabular}{cccccc} 
\toprule
\textbf{K} & \textbf{t} & \textbf{best classical} & \textbf{best quantum} & \textbf{separation} & \textbf{GD}\\
\midrule
1 & 32 & 0.61 & 0.72 & 0.11 & 2026.57\\
4 & 32 & 0.62 & 0.66 & 0.04 & 3417.23\\
6 & 32 & 0.59 & 0.80 & 0.21 & 2619.80\\
1 & 0.5 & 0.88 & 0.86 & -0.02 & 7.73\\
4 & 0.5 & 0.88 & 0.88 & 0.00 & 118.74\\
6 & 0.5 & 0.87 & 0.90 & 0.03 & 8.34\\
\bottomrule
\end{tabular}
\caption{List connecting the difference between the best classical and the best quantum accuracy (=separation) with the GD. The columns $K$ and $t$ indicate the hyperparameter values for which dataset 6 was relabeled. The other parameters were set to $basis = distance$, $T=9$, $K=6$, $\gamma = 1$ and $\alpha_k = 1$ (\cref{eq:rdm_inner} and \cref{eq:rdm_distance})}
\label{tab:rela_alpha_1}
\end{table}
\end{appendices}
\end{document}